\documentclass{aastex63}

\def\rr#1{{\protect{#1}}}
\usepackage[utf8]{inputenc}
\usepackage{tabularx}
\usepackage{float}
\usepackage{latexsym}
\def\ion#1#2{#1 \textsc{#2}}

\received{September 28, 2020}
\accepted{November 15, 2020}
\submitjournal{APJ}

\shorttitle{Spectral Implications of Atomic Uncertainties in Optically-thin Hot Plasmas}
\shortauthors{Heuer et al.}
\date{\today}

\begin{document}

\title{Spectral Implications of Atomic Uncertainties in Optically-thin Hot Plasmas}

\author[0000-0002-6395-4528]{Keri Heuer}
\affil{Department of Physics, Drexel University, \\ 3141 Chestnut St, Philadelphia, PA 19104, USA} 
\affil{Center for Astrophysics | Harvard \& Smithsonian, \\
  60 Garden Street, Cambridge, MA 02138, USA}

\author[0000-0003-3462-8886]{Adam R. Foster}
\affil{Center for Astrophysics | Harvard \& Smithsonian, \\
  60 Garden Street, Cambridge, MA 02138, USA}

\author[0000-0003-4284-4167]{Randall Smith}
\affil{Center for Astrophysics | Harvard \& Smithsonian, \\
  60 Garden Street, Cambridge, MA 02138, USA}

\correspondingauthor{Keri Heuer}
\email{kh3286@drexel.edu}

\begin{abstract}
Two new high-resolution X-ray spectroscopy missions, XRISM and Athena, will observe deeper and with higher X-ray resolution than ever before possible. Interpreting these new X-ray spectra will require understanding the impact that uncertainties on fundamental atomic quantities such as collisional cross sections, transition rates, and wavelengths have on spectral models. As millions of values are required to generate even a simple model of an optically-thin hot plasma, most such rates exist only as theoretical calculations. We have developed methods to estimate the uncertainty in the final spectral calculations based on published experimental data and plausible approximations to the uncertainties in the underlying atomic data. We present an extension to the \texttt{pyatomdb} code that implements these methods and investigate the sensitivity of selected strong diagnostic lines in the X-ray bandpass (0.3-12 keV).
\end{abstract}


\section{Introduction}
X-ray-emitting optically-thin collisional plasmas exist throughout astrophysics on a range of scales, from stellar coronae to galaxy clusters. The X-ray emission arises from a range of processes, most prominently including bremsstrahlung, ion line emission, and radiative recombination. Diagnosing basic underlying plasma parameters does not always require high-resolution data. At electron temperatures above 2 keV, bremsstrahlung radiation dominates the emission from astrophysical plasmas with a broad and well-understood continuum. Even at lower temperatures, early X-ray grating spectrometers were, despite limited resolution, able to make accurate inferences about the underlying system. \citet{Holt1979X-rayCapella.} used Einstein Solid-State Spectrometer ($R\sim 140$) observations of Capella to correctly conclude the corona has at least two components between $6-24\times10^6$K and to detect Mg, Si, S, and Fe.  These results were primarily based on detecting a continuum that could not be fit with a single bremsstrahlung component along with models for the strong lines from He-like ions of Mg, Si, S, and Fe.

More detailed analyses with higher-resolution spectrometers, however, require both more and better atomic data.  In some cases, these data can be directly tied to laboratory measurements. Charge state distributions (CSDs\footnote{We use CSD, rather than ionization balance because the latter implies the system is in some type of equilibrium, which may or may not be true.}) in hot plasmas, for example, depend upon a modest number of ionization and recombination rate coefficients.  Many different collections have been used throughout the history of X-ray astronomy. Early compilations such as \citet{Jordan1969TheNickel} and \citet{Raymond1977SoftPlasma.} were updated \citep{Sutherland1993CoolingPlasmas,Mazzotta1998IonizationNI} as newer data became available (e.g. dielectronic recombination rates) or it was realized that processes such as excitation-ionization were dramatically underestimated.  The process has continued \citep{Bryans2006CollisionalIons,Bryans2009ACoefficients}, and these updates have at times significantly changed the conclusions observers draw.  The update of the AtomDB atomic database \citet{Smith2001CollisionalIonsb, Foster2012UpdatedSpectroscopy} from the \citet{Mazzotta1998IonizationNI} CSD to the calculations of \citet{Bryans2006CollisionalIons,Bryans2009ACoefficients} showed that the ISM of the elliptical galaxy NGC 4649 could be explained with a single temperature, rather than requiring a physically-unmotivated two component model \citep{Loewenstein2012An4649}.  Nonetheless, for CSD calculations at least, the process seems to be converging. The most recent compilation of ionization rates by \citet{Urdampilleta2017X-rayZn} showed ``differences less than 10-20\% for more than 85\% of the elements,'' a substantial improvement from previous updates, although the prospect of including a $\sim 15\%$\ error into collisional spectral models suggests work remains to be done.
 
The situation with line emission remains less well understood. High-resolution X-ray grating observations of stars with Chandra and XMM-Newton have revealed some problems with the existing collisional plasma models \citep[\protect{\it e.g.}][]{Huenemoerder2003TheLacertae}.  However, stellar coronae have complex temperature distributions and densities high enough to impact line emission, making it challenging to determine if a poor fit is due to the atomic model or the physical parameters used.  Recently, the Hitomi spectra of the Perseus cluster \citep{HitomiCollaboration2016TheCluster} highlighted the need to explore the sensitivity of measurements to atomic line emission data. Unlike a highly-dynamic magnetized corona, the diffuse emission from a galaxy cluster like Perseus is relatively simple to model, and showed a number of flaws in both the AtomDB and SPEX \citep{Kaastra1996SPEX:Spectra.} models used to fit it.  
The Hitomi Collaboration made an initial effort in their atomic data paper \citep[][hereafter ``\rr{HADP2018}'']{HitomiCollaboration2018AtomicHitomi} to estimate uncertainties in fluxes from lines of H- and He-like ions using disagreements between SPEX 3.03, ATOMDB 3.0.8 and other databases. \rr{HADP2018} thus demonstrated uncertainty estimates were both feasible and useful but focused only on a few lines. 

Here we expand upon this approach to consider uncertainties in the CSD as well as emission lines, based on the wavelengths, rates, and cross sections in AtomDB database. Using \texttt{variableapec} 1.0.2\footnote{http://github.com/AtomDB/variableapec}, we perturb values such as collisional cross sections and recalculate the CSD and level populations. This allows an encapsulation of an overall uncertainty into a single quantity, such as the line emissivity for a collisional plasma ($\epsilon$(n,T) in units of photons cm$^{3}$ s$^{-1}$). While the size of our perturbations are motivated by existing measurements, we emphasize that these calculations are at best estimates. In particular, we have no method to include correlated errors, such as would occur from changes to a structure calculation of an ion's electronic energy levels \citep[\protect{\it e.g.}][]{Loch2013TheModels}. Nonetheless, we hope that the results can be used by observers to estimate potential uncertainties in their model parameters from the underlying atomic data. 

\section{CSD Uncertainty Calculation}

\subsection{Method}
We use related but distinct methods to calculate uncertainties in the CSD and the line emissivities. The rationale for this separation is a combination of issues. First, for a given element $Z$\ and electron temperature $T_e$, the CSD calculation involves $\sim Z$\ ionization and recombination rates. Depending upon the ion and the level of detail in the calculation, metastable levels and multiple-ionization rates may be included, but in any event there is limited amount of atomic data required.  This cannot be said of line emission calculations, which rely upon theoretical calculations of atomic structure with hundreds to thousands of energy levels per ion with collisional and radiative transitions potentially allowed between any of these. Secondly, the CSD rates have some of the best experimental measurements of any values required for collisional plasma models, while electronic excitation cross sections and radiative transitions are dominated by theoretical calculations and have at best spot measurements.  

The underlying atomic processes driving the CSD are measured experimentally and calculated theoretically as cross sections as a function of electron energy. These process include direct ionization, excitation-autoionization, radiative recombination, and dielectronic recombination \citep[see \protect{\it e.g.}][]{Smith2014ChapterPlasmas}. For a thermal plasma model, theoretical calculations are typically compared to experimental measurements (and perhaps adjusted) and then integrated over a Boltzmann distribution to derive the ionization or recombination rate as a function of temperature.

\citet{Bryans2006CollisionalIons} updated the largely theoretical calculation of ionization rates by \citet{Dere2007IonizationZinc} and recombination rates by \citet{Badnell2006RadiativePlasmas}. \citet{Urdampilleta2017X-rayZn} has since provided another update of ionization rates with an estimate of uncertainty in experimental cross sections. The various rate errors from different experimental data sets can result in vast differences of CSD for an otherwise identical plasma. These updates prompt a robust study of the effect of atomic uncertainties on the CSD, which up until now has yet to be investigated in depth. 

To estimate a reasonable systematic uncertainty, we conducted a thorough literature search of experiments measuring energy-dependent cross sections $\sigma(E)$\ or Maxwellian-averaged rate coefficients $\Upsilon(T_e)$\footnote{As both types of data are regularly provided, we refer to ionization or recombination `values' to indicate when either is possible.}.  Tables \ref{tab:ionization} and \ref{tab: recombination} list systematic uncertainties for ionization and recombination values respectively from the most recent experiments for highly charged Fe and O ions. These also list the type of experiment and for Table~\ref{tab:ionization} notes if the result is an energy-dependent electron-impact ionization (EII) cross section or a Maxwellian-averaged ionization (I) rate coefficient. In most cases papers listed only a single systematic error, but in cases where multiple errors were listed, {\it e.g.}\ at different energies, the value \rr{closest to the threshold for ionization} was selected.
Based on this catalog of \rr{multiple} systematic uncertainties \rr{for each ion}, we use the minimum error found for each ion's ionization and recombination rate coefficient for our thermal CSD calculations \rr{using the collisional equilibrium model}. The minimum was chosen in order to not overestimate the effects of systematic errors. For our other calculations looking at the \rr{sensitivity of ionic concentration} to error magnitude independent of this tabulation, we considered a range of perturbations, typically from 5-40\%, and use the same change for both ionization and recombination rate coefficients. \rr{Our calculated changes in ionic concentration are thus the minimum due to systematic uncertainties. In cases where multiple systematic errors are listed at different energies, the resulting change in concentration can therefore be larger.}

We also note any published comparisons \rr{to rates (``R'', either cross sections or rate coefficients) in} theoretical data sets since it is informative when updated experimental results are compared to commonly-used theoretical values. $\Delta$R values \rr{are defined as $(R_{exp} - R_{th})/R_{exp}$\ where these values are available; listed values from the literature ({\it e.g.}, from text such as ``Our rates agree to theory within 10\%'') are assumed to mean the same.} These are listed only for the most recent theoretical data set if a reference has compared their experimental data to multiple sources; see respective references for comparisons with older theories. Where references have calculated their own theoretical values for comparison with experimentally-derived rate coefficients, we list the source for $\Delta$R as self. 

\begin{figure}
    \centering
    \includegraphics[width=8cm]{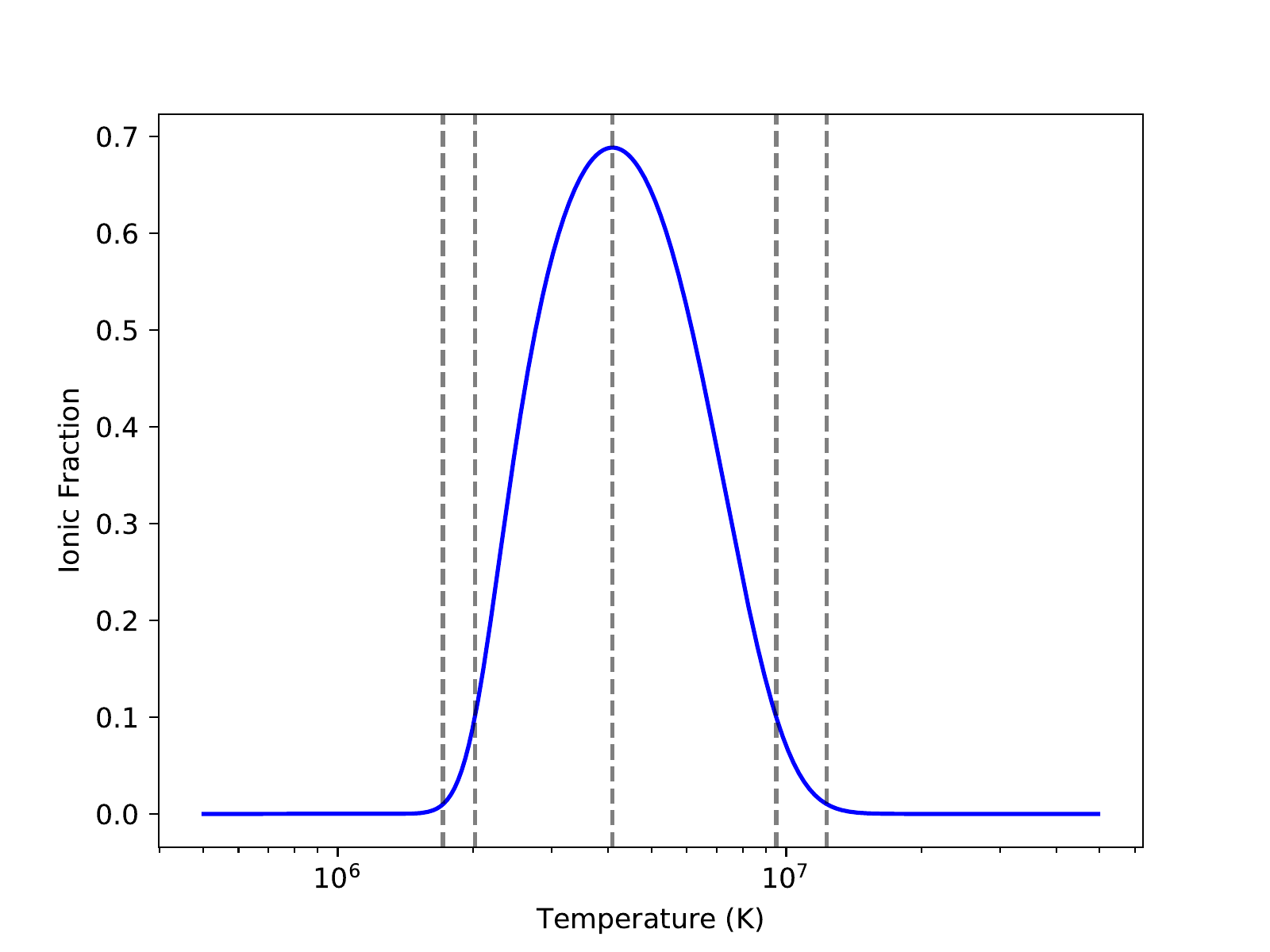}
    \caption{Five temperatures of interest for \ion{Fe}{xvii} indicated by horizontal dashed lines on the low-temperature (`rising') side at 1\% and 10\% \rr{ionic concentration}, the peak temperature, and the high-temperature (`falling') side at 1\% and 10\%.}
    \label{fig:5_temps}
\end{figure}

To understand the effect of atomic uncertainties on the CSD, we first perturb a single rate coefficient \rr{by the same amount at all temperatures} and calculate new ion populations for temperatures where the fractional \rr{ionic concentration} is greater than $10^{-4}$. CSD curves generally appear to be Gaussian, although the wings and details depend upon the specific ion. To quantify the impact of the uncertainty on the inferred temperature of the plasma, we considered the fractional change \rr{in ionic concentration} at five temperatures: where the ion fraction is rising as a function of temperature to 1\%, 10\%, and then peak, and as it drops back down again to 10\% and 1\% [see Figure \ref{fig:5_temps}]. Comparing the \rr{rate of change in ionic concentration as a function of fractional change, defined as half the change in emissivity calculated from the minimum and maximum CSD, on a given rate coefficient at these five temperatures} indicates the temperatures where an ion is most affected by uncertainties \rr{[see Figure \ref{fig:slope}]}. We repeat this process for various levels of uncertainties to compare the resulting CSD and sensitivities to perturbation magnitude. 

Using random errors selected from a flat distribution (chosen as these are systematic, not statistical, errors), we then jointly perturb all rate coefficients for an element and use Monte Carlo calculations to derive the range of potential population values for each ion over a range of plasma temperatures. 
For ions where no errors have been published, we employ an average of the errors across all ions in Tables \ref{tab:ionization} and \ref{tab: recombination}. Errors ($\sigma$) are selected from the minimum error on the rate coefficient we employ from Tables \ref{tab:ionization} and \ref{tab: recombination}. We then compare the non-perturbed ion population over Monte Carlo approximations with $\pm\sigma$ error bars to derive a range of potential population values for an ion over a range of plasma temperatures.

\startlongtable
\begin{deluxetable}{ccccccc}
    \tablecaption{Systematic uncertainties for experimentally measured ionization rate coefficients. \rr{$\Delta$R is defined as the published  comparison to the theoretical  data  set listed as the ``comparison''. If the reference calculated their own theoretical values, the comparison source is listed as ``self''.}\label{tab:ionization}}

    \tablehead{
     \colhead{Ion} & \colhead{Error} & \colhead{Type\tablenotemark{a}} & \colhead{Method\tablenotemark{b}} & \colhead{Reference} &  \colhead{ $\Delta$R} & \colhead{Comparison}}

    \startdata
    O 7+ & 10\% & EII & cb & \citet{Aichele1998ElectronO7+} & & \\
    O 6+ & 29\% & EII & ps & \citet{Rowan1979Electron-impactOxygen} & 35\% & \citet{1974MNRAS.169..663S}\\
     & 20\% & EII & ps & \citet{Kallne1977MeasurementsIons} & & \\
    O 5+ & 11\% & EII & cb & \citet{Crandall1986Electron-impactIons} & 50\% & \citet{Jakubowicz1981ElectronIons} \\
     & 8\% & EII & cb & \citet{Rinn1987Electron-impactMeasurementsb}& 9\% &  \citet{Pindzola1986Electron-impactSequence}\\
     & 8\% & EII & cb & \citet{Muller2000AutoionizingIons} & & \\
    O 4+ & 8\% & EII & cb & \citet{Fogle2008Electron-ImpactV} & 17\% & \citet{Dere2007IonizationZinc} \\
     & 7\% & EII & cb & \citet{Loch2003Electron-impact4} & &\\
     & 6\% & EII & cb & \citet{Falk1983Measured+} & 5\% & \citet{1981PhRvA..24.1278Y}\\
    Fe 25+ & 15\% & EII & e & \cite{ORourke2001Electron-impactIons} & & \\
    Fe 23+ & 13\% & EII & e & \citet{Wong1993Electron-impactFe23+} & 4\% & \citet{Zhang1990RapidIons} \\
    Fe 21+ & 40\% & I & t & \citet{Wang1988MeasurementPlasmas} & 33\% &  \citet{1977JPhB...10.2229G}\\
    Fe 20+ & 40\% & I & t & \citet{Wang1988MeasurementPlasmas} & 0\% & \citet{1977JPhB...10.2229G} \\
    Fe 19+ & 40\% & I & t & \citet{Wang1988MeasurementPlasmas} & 64\% & \citet{1977JPhB...10.2229G} \\
    Fe 18+ & 40\% & I & t & \citet{Wang1988MeasurementPlasmas} & 20\% & \citet{1977JPhB...10.2229G} \\
    Fe 17+ & 40\% & I & t & \citet{Wang1988MeasurementPlasmas} & & \\
     & 16\% & EII & cb & \citet{Hahn2013StorageAstrophysics} & 4-17\% & \citet{Dere2007IonizationZinc} \\
     & 15\% & EII & cb & \citet{Hahn2014ElectronIons} & & \\
    Fe 16+ & 15\% & EII & cb & \citet{Hahn2014ElectronIons} & 10-15\% & \citet{Dere2007IonizationZinc} \\
     & 40\% & I & t & \citet{Wang1988MeasurementPlasmas} & & \\
      & 16\% & EII & cb & \citet{Hahn2013StorageAstrophysics} & 1-19\% & \citet{Dere2007IonizationZinc} \\
    Fe 15+ & 20\% & EII & cb & \citet{Hahn2014ElectronIons} & & \\
     & 40\% & I & t & \citet{Wang1988MeasurementPlasmas}  & 15\% & \citet{Pindzola1986Electron-impactSequence} \\
     & 20\% & EII & cb & \citet{Muller1999PlasmaRings} & 20\% & \citet{Arnaud1992IronEquilibrium} \\ 
    Fe 14+ & 26\% & EII & cb & \cite{Bernhardt2014AbsoluteRing} & 35-70\% & \citet{Arnaud1992IronEquilibrium} \\
    Fe 13+ & 16\% & EII & cb & \citet{Hahn2014ElectronIons} & 35\% & \citet{Gregory1987Electron-impact+} \\
     & 6\% & EII & cb & \citet{Hahn2011StorageFe13+} & 15\% & \citet{Dere2007IonizationZinc} \\
      & 16\% & EII & cb & \citet{Hahn2013StorageAstrophysics} & 35\% & \citet{Arnaud1992IronEquilibrium}\\
    Fe 12+ & 6\% & EII & cb & \citet{Hahn2011Storage+}& 15\% & \citet{Dere2007IonizationZinc} \\
    Fe 11+ & 16\% & EII & cb & \citet{Hahn2011StorageFe13+} & 15\%, 30\% & self, \citet{Arnaud1992IronEquilibrium} \\
     & 15\% & EII & cb & \citet{Hahn2014ElectronIons} & & \\
     & 14\% & EII & cb & \citet{Gregory1987Electron-impact+} & 30\% &  \citet{Pindzola1986Electron-impactSequence} \\
    Fe 10+ & 9\% & EII & cb & \citet{2012ApJ...760...80H} & 19\% & \citet{Dere2007IonizationZinc} \\
    Fe 9+  & 15\% & EII & cb & \citet{Hahn2014ElectronIons} & & \\
     & 9\% & EII & cb & \citet{2012ApJ...760...80H} & 16\% & \citet{Dere2007IonizationZinc} \\
    Fe 8+ & 16\% & EII & cb & \citet{Hahn2016StorageFe8+} & 15-40\% & \citet{Dere2007IonizationZinc} \\
    Fe 7+ & 12\% & EII & cb & \citet{2015ApJ...813...16H} & 10\% & \citet{Dere2007IonizationZinc} \\
    \enddata
    \tablenotetext{a}{Electron-impact ionization (EII) or effective ionization rate (I).}
    \tablenotetext{b}{Crossed-beam apparatus (cb), plasma spectroscopy/$\theta$-pinch (ps), electron beam ion trap (e), tokamak (t).}
\end{deluxetable}

\startlongtable
\begin{deluxetable}{ccccccc}
    \tablecaption{Systematic uncertainties for experimental recombination rate coefficients. \rr{$\Delta$R is defined as in Table 1.}\label{tab: recombination}}
    \tablehead{
     \colhead{Ion} & \colhead{Error} & \colhead{Type\tablenotemark{a}} & \colhead{Method\tablenotemark{b}} & \colhead{Reference} & \colhead{$\Delta$R} & \colhead{Comparison}}
    
    \startdata
    O 7+ & 25\% & DR & cb & \citet{1990PhRvL..64..737K}& 20\% & \citet{Bell:1981vja} \\
      & 18\% & RR & cb & \citet{1990JPhB...23.3167A} & & \\
    O 6+ & 20\% & DR & cb & \citet{1990PhRvA..41.1293A} & & \\
    O 5+ & 11\% & DR & cb & \citet{Bohm2002MeasurementN} & 15-25\% &  self \\
     & 20\% & DR & cb & \citet{1990PhRvA..41.1293A} & & \\
      & 35\% & DR & cb & \citet{Dittner1987DielectronicO5+} & 30-60\% & \citet{1986PhRvA..34.3668P}\\
     & 18\% & RR & cb & \citet{1990JPhB...23.3167A} & & \\
    O 4+ & 30\% & DR & cb & \citet{Dittner1987DielectronicF5+} & & \\
    Fe 24+  & 20\% & DR & cb & \citet{Gwinner2001DielectronicIons} & & \\
    Fe 22+ & 20\% & DR & cb & \citet{Savin2006DielectronicCalculations} & 20\% & self \\
     & 20\% & DR & cb & \citet{Schippers2010DielectronicExperiments} & & \\
      & 20\% & DR & cb & \citet{Gwinner2001DielectronicIons} & & \\
    Fe 21+ & 20\% & DR & cb & \citet{Schippers2010DielectronicExperiments} & & \\
     & 20\% & DR & cb & \citet{2013HEAD...1312201S} & & \\ 
       & 20\% & DR & cb & \citet{Gwinner2001DielectronicIons} & & \\
    Fe 20+ & 20\% & DR & cb & \citet{Schippers2010DielectronicExperiments} & & \\
     & 20\% & DR & cb & \citet{2013HEAD...1312201S} & & \\
       & 20\% & DR & cb & \citet{Gwinner2001DielectronicIons} & & \\
    Fe 19+ & 20\% & DR & cb & \citet{Savin2002DielectronicCalculations} & 15\%, 30\% & self, \citet{Zatsarinny2004DRCarbon} \tablenotemark{c}\\ 
     & 20\% & DR & cb & \citet{Schippers2010DielectronicExperiments} & & \\ 
      & 20\% & DR & cb & \citet{Gwinner2001DielectronicIons} & & \\
    Fe 18+ & 20\% & DR & cb & \citet{2002nla..work...83S} & 20\%, 40\% & self, \citet{Wang1988MeasurementPlasmas} \\
     & 20\% & DR & cb & \citet{Schippers2010DielectronicExperiments} & & \\
     & 20\% & DR & cb & \citet{Gwinner2001DielectronicIons} & 160\% & \citet{Savin1999DielectronicXIX} \\
    Fe 17+ & 20\% & DR & cb & \citet{1997APS..APR.D1557S} & 200\%, 30\% & \citet{Chen1988DielectronicCalculations}, self \\
     & 20\% & DR & cb & \citet{Schippers2010DielectronicExperiments} & 30\% & self, \citet{1999ApJS..123..687S} \\
      & 20\% & DR & cb & \citet{Gwinner2001DielectronicIons} & 200\% & \citet{Savin1999DielectronicXIX}\\
    Fe 16+ & 22\% & DR & cb &\citet{Schippers2010DielectronicExperiments} & & \\
     & 20\% & DR & cb &\citet{Schmidt2009ExperimentalIron} & 14\% & \citet{Zatsarinny2004DRNeon}\\
    Fe 15+ & 20\% & DR & cb &\citet{Linkemann1995ElectronIons} & 200\% &\cite{Chen1990ContributionsFe15+} \\
     & 20\% & DR & cb & \citet{Schippers2010DielectronicExperiments} & & \\
    Fe 14+ & 26\% & DR & cb & \citet{Bernhardt2014AbsoluteRing} & 26\% & \citet{Altun2007DRMagnesium} \\ 
    & 29\% & DR & cb & \citet{Lukic2007DielectronicCalculations} & 19-37\%, 300\% & self, \citet{Arnaud1992IronEquilibrium} \\
     & 29\% & DR & cb & \citet{Schippers2010DielectronicExperiments} & & \\
    Fe 13+ & 18\% & DR & cb & \citet{Schmidt2006Electron-IonXIII}& 21\% & \citet{Arnaud1992IronEquilibrium}\\
     & 29\% & DR & cb & \citet{Schippers2010DielectronicExperiments} & & \\
    Fe 12+ & 18\% & DR & cb & \citet{Hahn2014ElectronIons} & 30\% & \citet{Badnell2006RadiativePlasmas}\\
    Fe 11+ & 13\% & DR & cb & \citet{2012ApJ...753...57N} & 30\% & \citet{Badnell2006RadiativePlasmas} \\
    Fe 10+ & 30\% & DR & ps & \citet{Brooks1980MeasurementXi} &70\% & \citet{1977ApJ...211..605J}  \\
     & 25\% & DR & cb & \citet{Schippers2010DielectronicExperiments} & & \\
     & 25\% & DR & cb & \citet{Lestinsky2009Electron-IonCalculations} & 25\% & \citet{Badnell2006RadiativePlasmas}  \\
    Fe 9+ & 30\% & DR & ps & \citet{Brooks1980MeasurementXi} & 150\% &  \citet{1977ApJ...211..605J}  \\
     & 25\% & DR & cb & \citet{Schippers2010DielectronicExperiments} & & \\
     & 25\% & DR & cb & \citet{Lestinsky2009Electron-IonCalculations} & 25\% & \citet{Badnell2006RadiativePlasmas}  \\
    Fe 8+ & 35\% & DR & cb & \citet{2008AA...492..265S} & & \\
     & 30\% & DR & ps & \citet{Brooks1980MeasurementXi} & 13\% & \citet{1977ApJ...211..605J}  \\
      & 29\% & DR & cb & \citet{Schippers2010DielectronicExperiments} & & \\     
    Fe 7+ & 26\% & DR & cb & \citet{2008AA...492..265S} & 120\% & \citet{Arnaud1992IronEquilibrium}\\
     & 25\% & DR & cb & \citet{Schippers2010DielectronicExperiments} & & \\
    \enddata
    
    \tablenotetext{a}{Dielectronic recombination (DR) or radiative recombination (RR).}
    \tablenotetext{b}{Crossed-beam apparatus (cb), plasma spectroscopy/$\theta$-pinch (ps).}
    \tablenotetext{c}{30\% scatter quoted from \citet{Zatsarinny2004DRNeon}}

\end{deluxetable}

\subsection{Results}

For simplicity, we focus on results for oxygen and iron ions, but the \texttt{variableapec} tools developed for this project can be used for any atom or line in the AtomDB database with user-defined errors \rr{(see Appendix)}. We note that a complete review of all systematic errors in ionization and recombination rates for astrophysically abundant ions, along the lines of Tables \ref{tab:ionization} and \ref{tab: recombination} would be welcomed, but is beyond the scope of this initial effort. The primary challenge is to determine a reasonable set of estimated uncertainties for the underlying atomic data. The availability of experimental data depends upon a number of factors, including the ease of use (noble gases are easier to work with than metals) and their astrophysical abundance (a proxy for their importance). 

\begin{figure}[H]
\centering
\includegraphics[totalheight=7 cm]{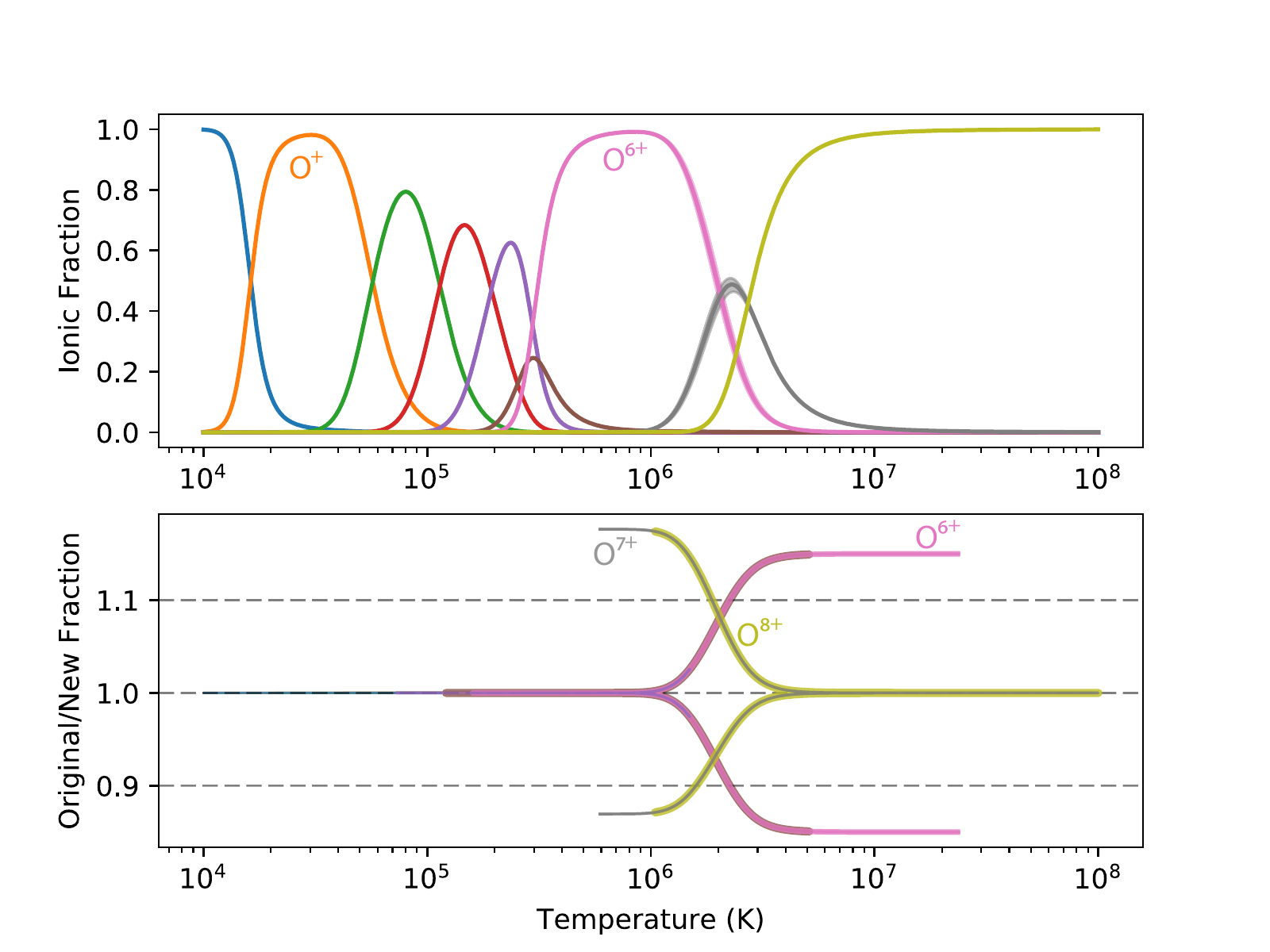}
\caption{[Top] Oxygen CSD as a function of temperature. [Bottom] \rr{Maximum} relative changes in the CSD after perturbing the O$^{6+}$ ionization rate by $\pm$15\%.}
\label{fig:csdo7}
\end{figure}

Changing even a single rate, such as the total ionization rate I(O$^{6+}\rightarrow$ O$^{7+}$), will potentially affect the population of a number of ions.  Figure~\ref{fig:csdo7} shows the impact of a $\pm15$\% change in this rate on the O$^{5+} - $O$^{8+}$\ ion populations. To avoid numerical issues, the calculations cut off at populations less than $10^{-5}$, resulting in the sharp bounds seen in the lower half of Figure~\ref{fig:csdo7}.  The relative impact of this one uncertainty is largest where the ion has a low \rr{ionic concentration}, showing that inferences made about temperature using an ion whose population is $<5\%$\ will potentially have large uncertainties, even though the apparent sensitivity is high due to a ion population that rapidly changes with temperature.  

\begin{figure}[H]
\begin{center}
\includegraphics[width=15 cm]{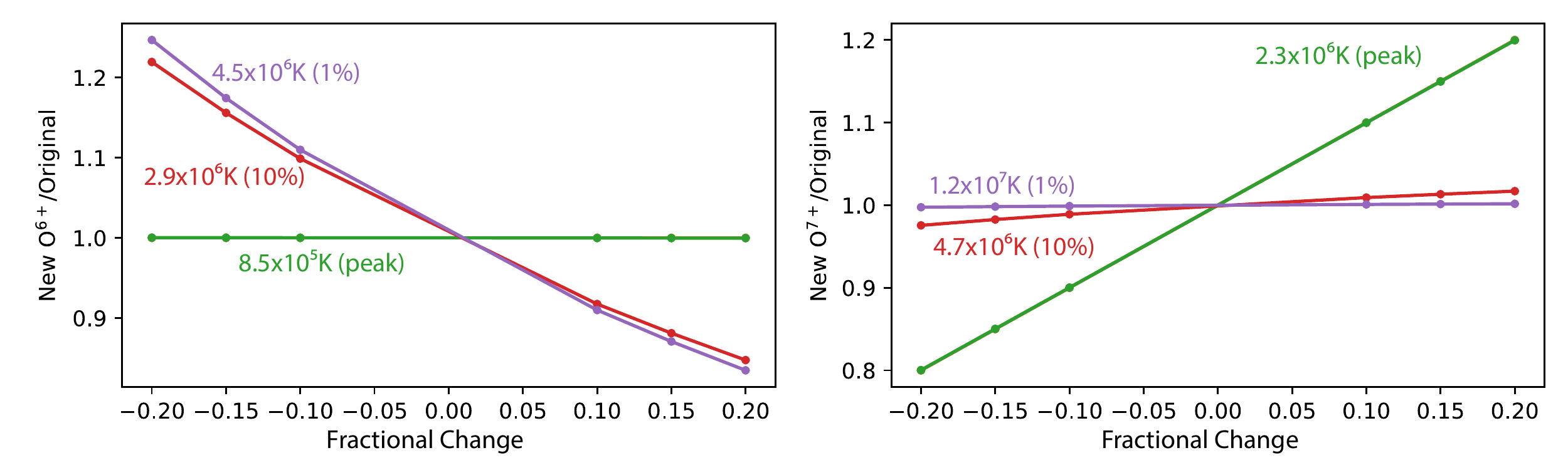}
\end{center}
\caption{[Left] Relative changes in O$^{6+}$ ion population at selected temperatures from a change in the O$^{6+}$\ ionization rate ranging from a -20\% decrease to a +20\% increase. No change is seen at temperatures at or below the peak of the \rr{ionic concentration (and so are not plotted)}, as there the \rr{ionic concentration} is dominated the balance between recombination with ionization from the lower ion O$^{5+}$. At temperatures above peak where the \rr{ionic concentration} has dropped to 10\% or 1\%, changes to the ionization rate directly affects the \rr{ionic concentration}. [Right] Same for O$^{7+}$ due to the perturbation in O$^{6+}$\ ionization rate.}
\label{fig:slope}
\end{figure}

This effect is explicitly shown in Figure~\ref{fig:slope}, which plots the changes in the \rr{ionic concentration} of O$^{6+}$ ([Left]) and O$^{7+}$ ([Right]) at the 5 temperatures shown in Figure~\ref{fig:5_temps} after modifying the O$^{6+}$\ ionization rate.  The lack of change at many temperatures shows that, in equilibrium, the exact value of each rate matters only over a small range of temperatures.  The ionization rate out of O$^{6+}$, for example, only matters if the plasma contains O$^{6+}$\ to ionize. In practice, the most sensitive range for ionization is usually close to the threshold, where calculations and measurements are most difficult.

\begin{figure}[H]
\includegraphics[width=8.5 cm]{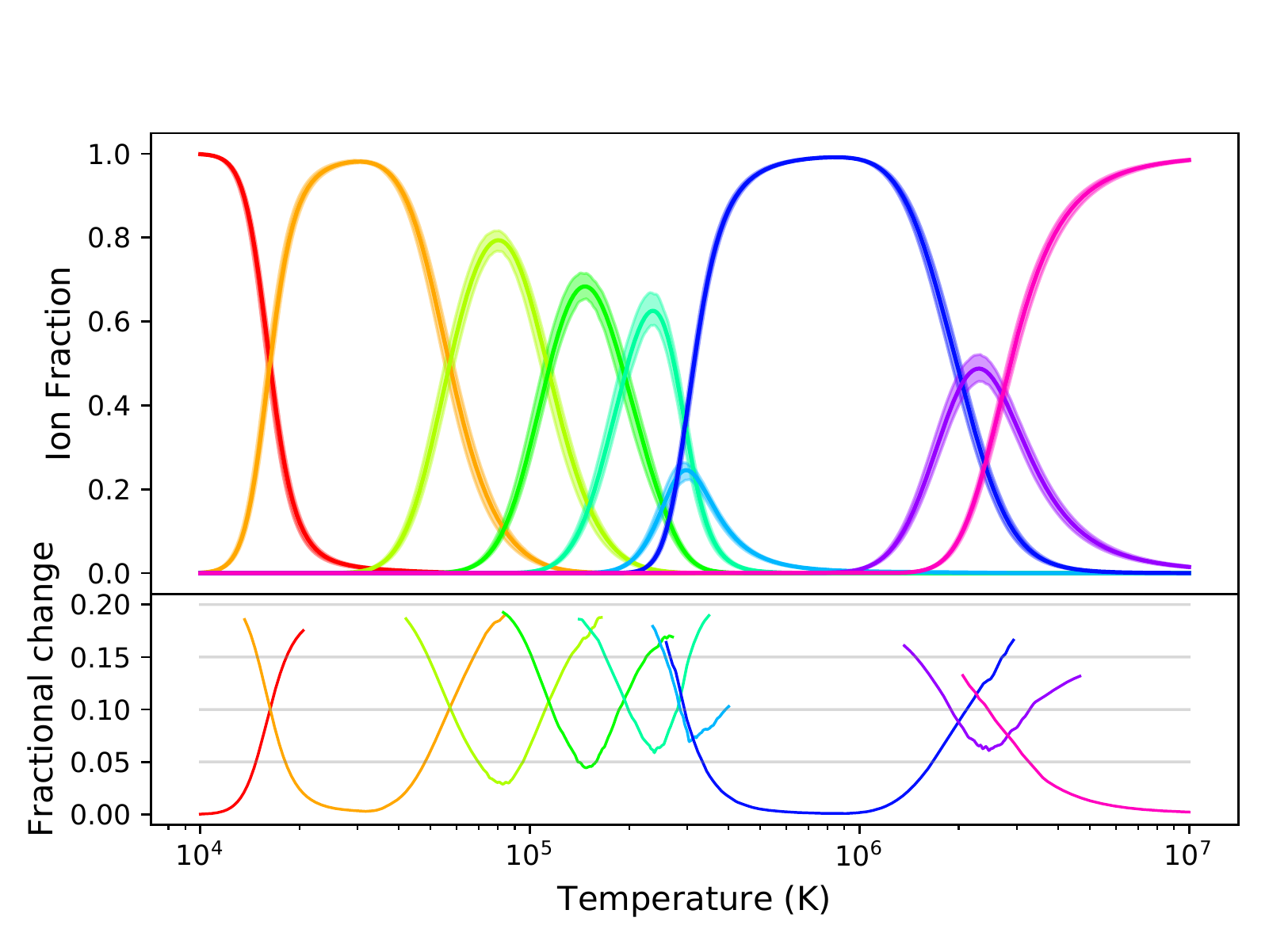}
\includegraphics[width=8.5 cm]{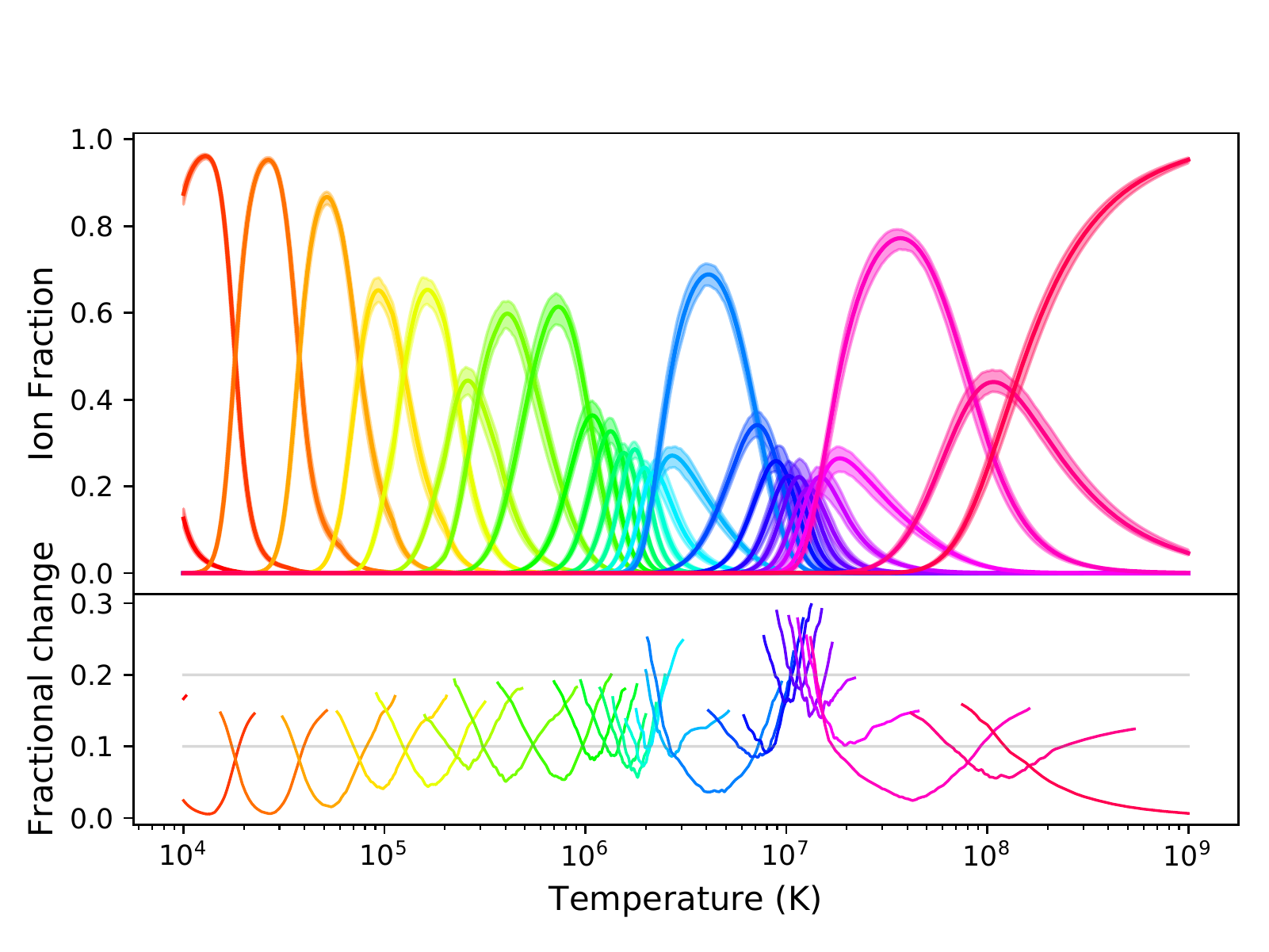}
\caption{[Left] CSD from perturbing all O rates with the minimum systematic errors listed in Tables \ref{tab:ionization} and \ref{tab: recombination} \rr{from 100 Monte Carlo runs}. [Right] Same, for Fe rates.}
\label{fig:new_csd}
\end{figure}

Figure~\ref{fig:new_csd} plots the perturbed CSD for all O and Fe atoms from Monte Carlo calculations and a flat distribution of errors using the systematic uncertainties for each ion in Tables \ref{tab:ionization} and \ref{tab: recombination}. For ions with multiple errors reported, we employ the smallest experimental error measured. For ions with no error listed, we average the errors from adjacent ions and for ions below O$^{4+}$\ and Fe$^{7+}$, we use the errors of O$^{4+}$\ and Fe$^{7+}$. The top panel of Figure \ref{fig:new_csd} shows the original CSD with the range of perturbed CSD, while the bottom panel shows the error in \rr{ionic concentration} for temperatures where the original \rr{ionic concentration} is greater than $10^{-3}$. Systematic uncertainties can result in an \rr{error on the ionic concentration} up to a factor of two for certain ions at temperatures where the ion is of low \rr{concentration}. `Transition' ions such as Li-like \ion{O}{vii} or Na-like \ion{Fe}{xvi} demonstrate strong sensitivity to temperature variations, making them often used for astrophysical analysis of spectra. Similarly the Fe L-shell ions (\ion{Fe}{xvii} - \ion{Fe}{xxiv}) provide a powerful diagnostic due to their dense and close-spacing in temperature. Unfortunately, this shows that CSD calculations for these ions also suffer disproportionately from systematic errors.

\begin{figure}[H]
\includegraphics[width=8.5 cm]{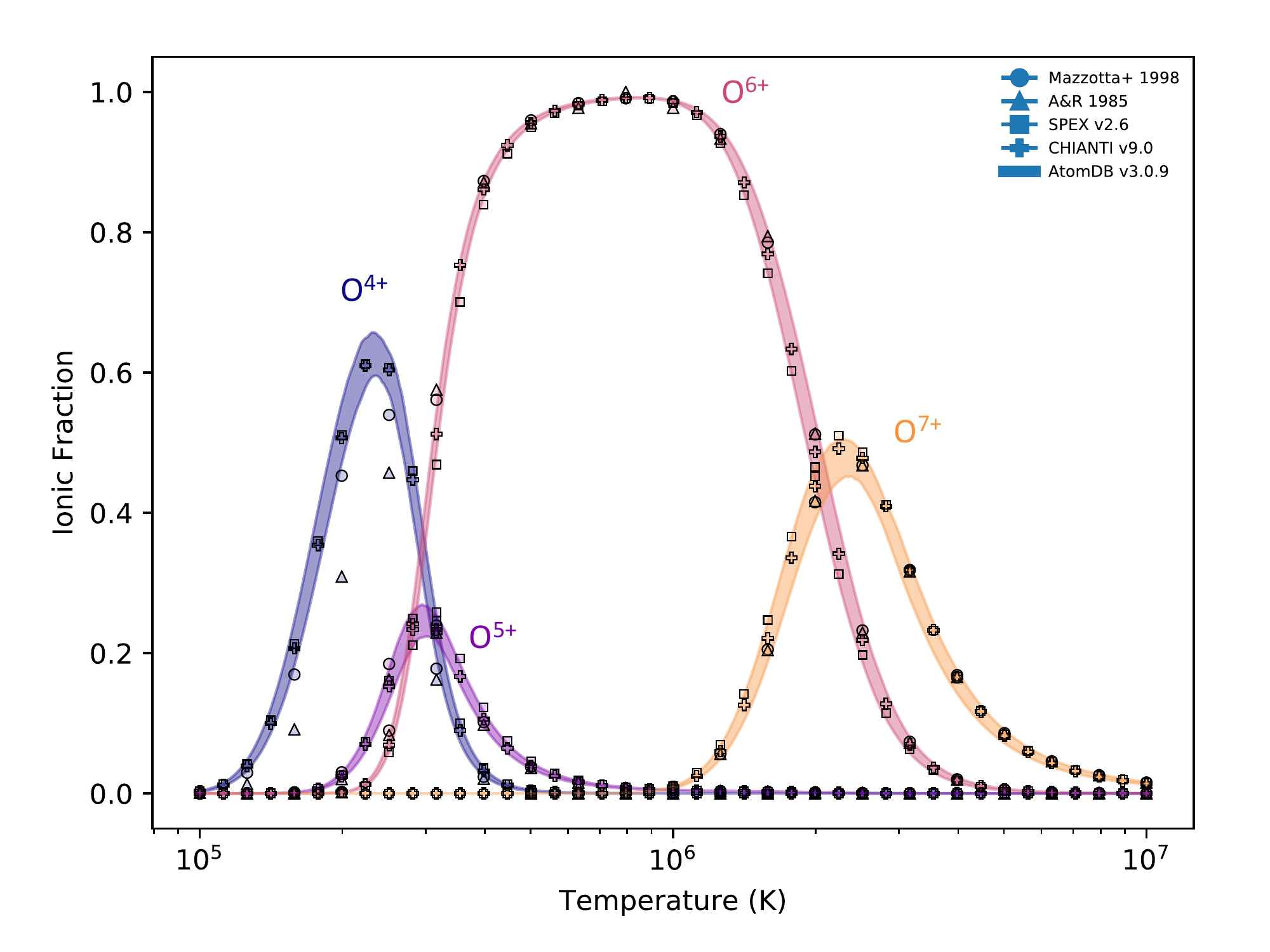}
\includegraphics[width=8.5 cm]{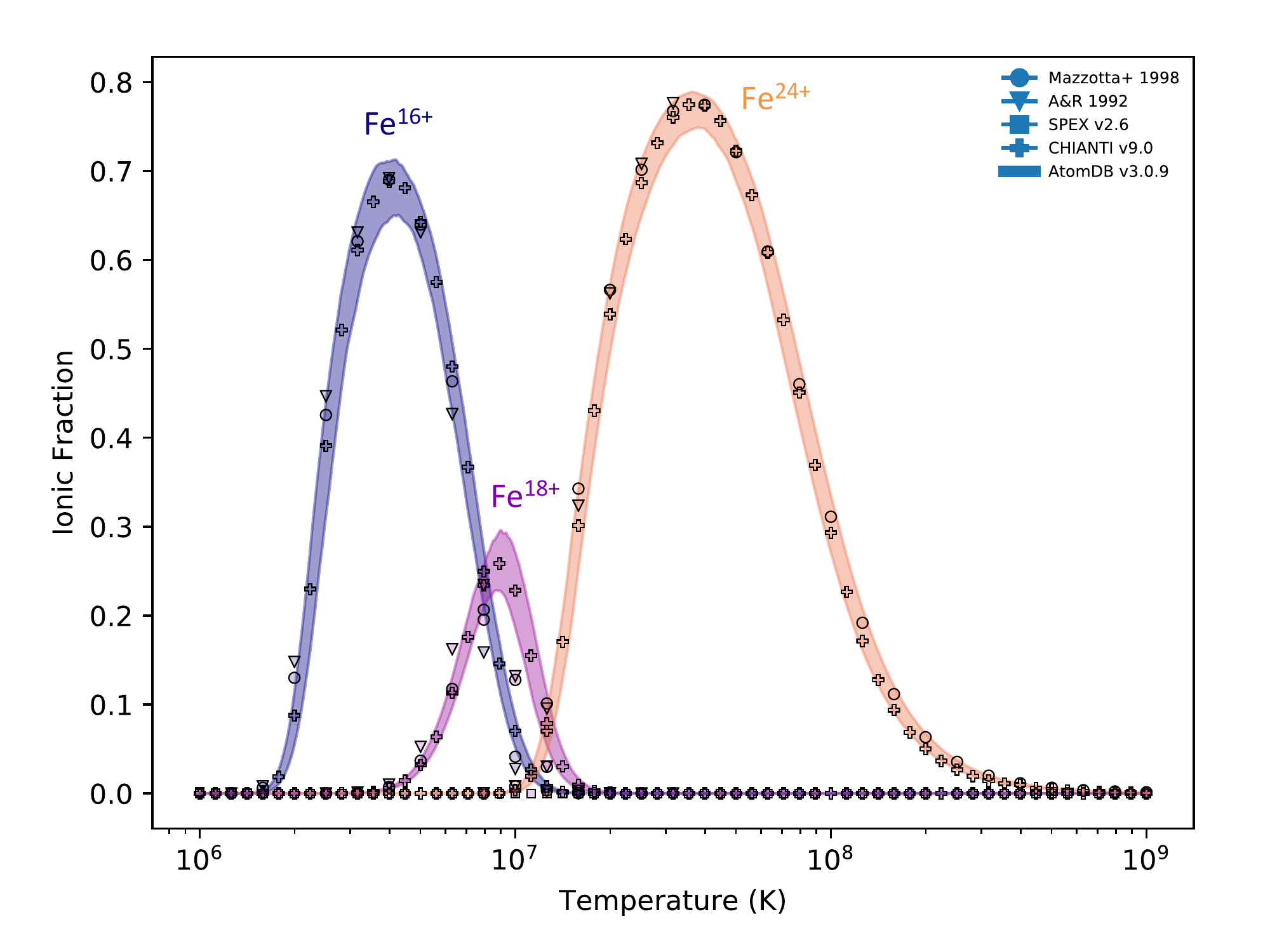}
\caption{Comparison of CSD for selected O [left] and Fe ions [right] from different data sets with \rr{the perturbed CSD range from minimum systematic uncertainties shown in Figure \ref{fig:new_csd}}.}
\label{fig:csd_comparison}
\end{figure}

\rr{HADP2018} considered a range of different atomic data compilations to estimate the underlying errors in the data.  Figure~\ref{fig:csd_comparison} shows the comparison of CSD for select O and Fe ions from commonly used compilations with the perturbed CSD range using the same errors employed for Figure \ref{fig:new_csd}, itself based on the \citet{Bryans2009ACoefficients} rates. The difference between the various data sets generally lies well within error range. O$^{4+}$\ has the greatest difference in \rr{ionic concentration}, with older data sets of \citet{Mazzotta1998IonizationNI} and \citet{Arnaud1985AnRates} lying outside the CSD calculated using the minimum experimental errors for these ions. Additionally, the temperature of peak \rr{ionic concentration} is slightly shifted for these two data sets. A visual comparison shows that recent theoretical data sets are in excellent agreement, in fact better than the systematic uncertainties in the underlying laboratory measurements. This suggests that future work needs to include more precise experimental data rather than just theoretical calculations.

\subsubsection{Ionic Concentration Calculations}

A common use of CSDs is to determine the relative abundance of an element based on emission or absorption from a single ion. \rr{More complex calculations will involve multiple ions, potentially including a combination of plasma temperatures. In the context of analyzing spectral data from such complex plasmas, the exact use of our results is beyond the scope of this paper and will require future work. For now, we can use our Monte Carlo technique to determine which individual ions are most sensitive to uncertainties even though we may not know the total errors on the ionization and recombination rates.} 

\begin{figure}[H]
\includegraphics[width=8.5 cm]{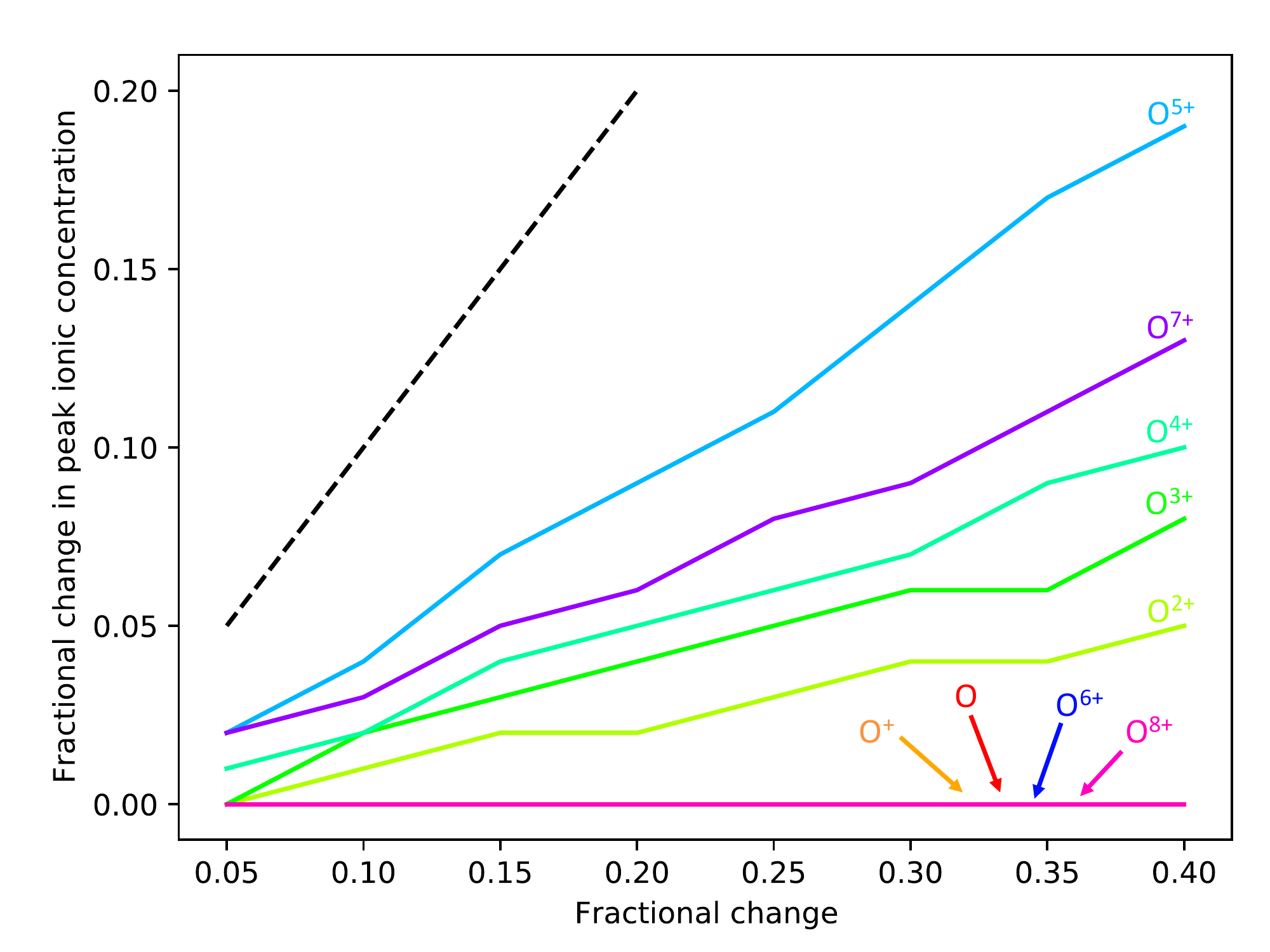}
\includegraphics[width=8.5 cm]{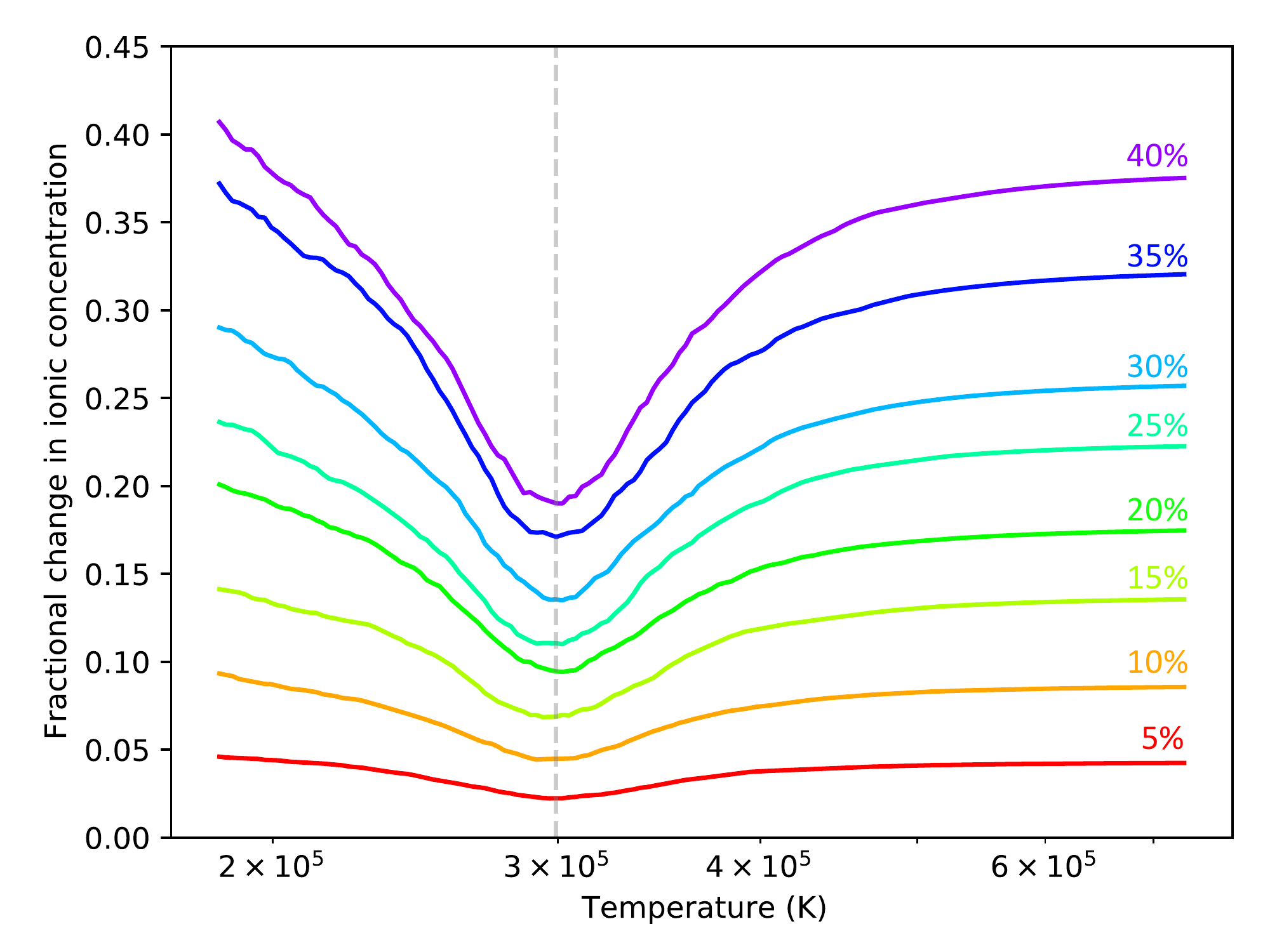}

\caption{[Left] Sensitivity of peak ionic concentration for all O ions from perturbing all rate coefficients as a function of the fractional change \rr{from 1000 Monte Carlo calculations}. The dashed black line shows a sample curve with unit slope, showing that the overall change in peak \rr{ionic concentration} for ions is lower than the errors on the total rates. [Right] Sensitivity of O$^{5+}$\ \rr{ionic concentration} over all \rr{temperatures where the ionic concentration is greater than 10e-3} to perturbing all rate coefficients between 5 and 40\% \rr{from the same 1000 Monte Carlo runs}. The vertical line is drawn at the temperature of the ion's \rr{peak concentration}.}
\label{fig:O_abund}
\end{figure}

As O$^{5+}$\ is both astrophysically relevant and the ion most sensitive to perturbations in rate coefficients, we investigate its peak \rr{ionic concentration} sensitivity by perturbing only its ionization or recombination rate coefficient and then perturbing all oxygen rate coefficients via Monte Carlo calculations. We find that the change in peak \rr{ionic concentration} is generally independent of which rate coefficient is varied, {\it i.e.} changing either the ionization or recombination rate coefficient yields the same fractional change in peak \rr{ionic concentration}. Experimental recombination rate coefficients, however, have a larger systematic error on average than ionization rate coefficients, so once both rate coefficients get perturbed, the \rr{change in peak ionic concentration} will be greater than the change from varying an individual rate coefficient. 

Figure~\ref{fig:O_abund} ([Left]) plots the fractional change in peak \rr{ionic concentration} for each oxygen ion as a function of the maximum assumed range used in the Monte Carlo calculations, taken to be the same for both ionization and recombination. Ions with large peak \rr{ionic concentrations} ({\it e.g.} neutral O or O$^{6+}$) naturally show small changes in this plot, since the peak is near unity in all cases. Less abundant transition ions such as  Li-like ion O$^{5+}$, however, have changes that amplify any systematic uncertainties in the atomic rates. For O$^{5+}$ specifically, a change of 20\% on rate coefficients can lead to \rr{less than a} 10\% change in \rr{ionic concentration} at the temperature O$^{5+}$ peaks at, while it results in changes $\sim$ 17\% at high temperatures and up to the full 20\% at low temperatures (Figure \ref{fig:O_abund} [Right]).

\subsubsection{Temperature Measurements via CSD}
\begin{figure}
    \centering
    \includegraphics[width=8.5cm]{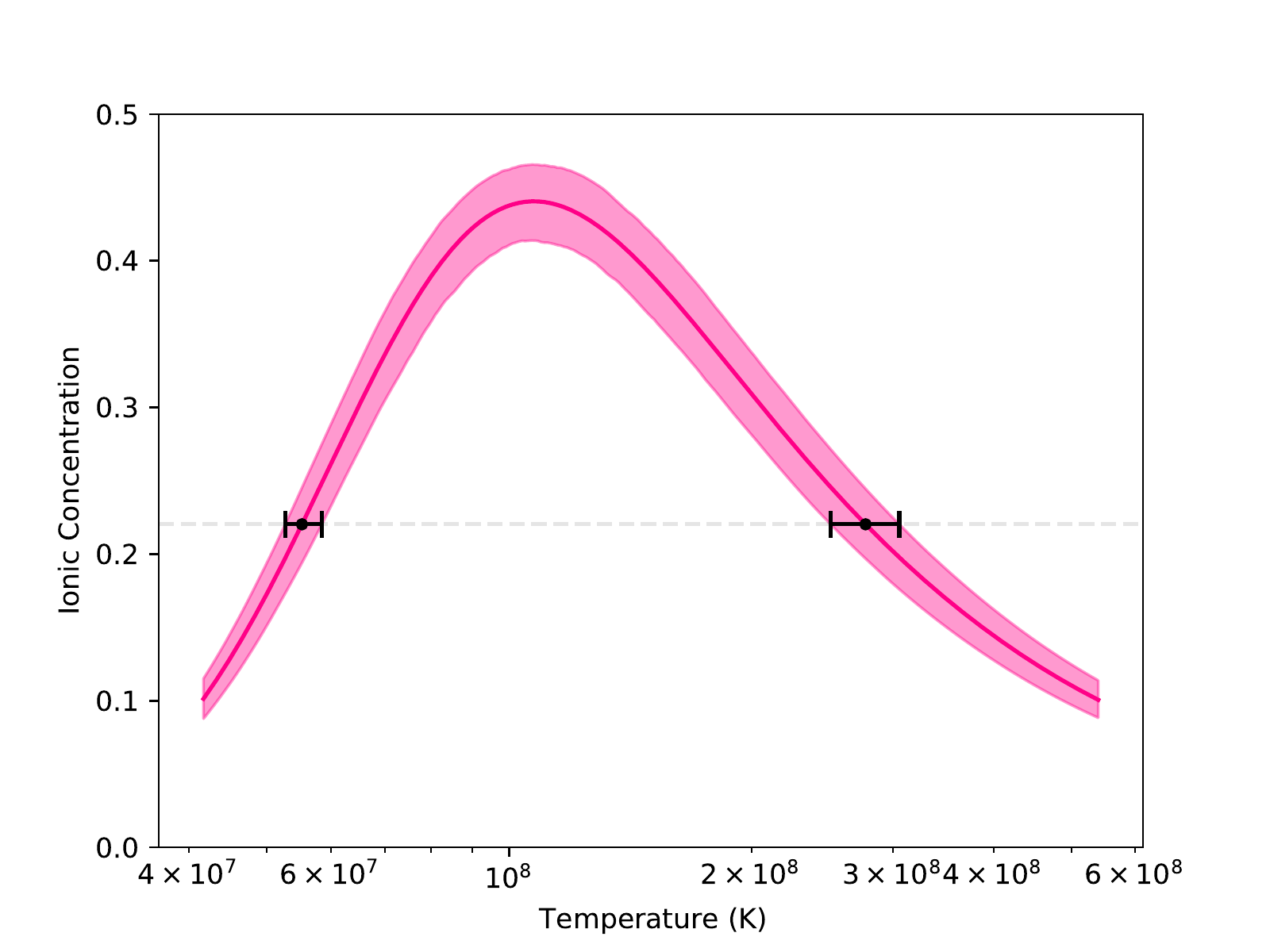}
    \caption{\rr{Temperature shifts (black error bars) for Fe $25^{+}$ from the unperturbed to perturbed CSD from Figure \ref{fig:new_csd} [right] at half peak ionic concentration (the horizontal dashed line). The temperature range plotted is where the ionic concentration is greater or equal to 10\%.}}
    \label{fig:temp_shift}
\end{figure}

Another common use of CSDs is to measure a plasma temperature by considering the ratio of lines from adjacent ions, such as \ion{O}{vii} and \ion{O}{viii}. Tables \ref{tab:O temp change} and \ref{tab:Fe temp change} list the implied temperature shifts in keV for O and Fe respectively from perturbing all rate coefficients with typical systematic errors. The temperature shifts listed are measured from the temperatures where the \rr{ionic concentration} in equilibrium is half of the peak value, both on the rising and falling sides [see Figure \ref{fig:temp_shift}]. These shifts can be treated as a type of systematic error on any temperature measurement made from these ions. \rr{We determined the temperature shifts from estimating the plasma temperature at half peak ionic concentration after performing Monte Carlo CSD calculations. As they were not determined through spectral fitting, these temperature shifts should be treated as a calculational exercise of estimating possible uncertainties in plasma temperature diagnostics due to typical systematics.}

Studying the fractional changes in peak \rr{ionic concentration} of ions over various magnitudes of perturbations to the rate coefficients is important for a quantitative suggestion for optimal error estimates. Optimizing temperature-dependent CSD errors is especially relevant when looking at specific temperature regions, such as the temperature ranges over which plasmas are formed through collisional ionization or photoionization. 

\begin{table}[H]
    \caption{Implied absolute temperature shifts (in keV) \rr{and relative temperature shifts (\%)} from perturbing all O ionization and recombination rates with the minimum systematic errors of Tables \ref{tab:ionization} and \ref{tab: recombination} at half peak \rr{ionic concentration} on the unperturbed rising ($T_{rise}$) and falling ($T_{fall}$) CSD curve.}
    \begin{tabular}{ccccccc}
    \hline
      \rr{Ion} & $T_{rise}$ & \multicolumn{2}{c}{\rr{Rising Shift}} & $T_{fall}$ & \multicolumn{2}{c}{\rr{Falling Shift}} \\ 
                & keV & keV & \rr{\%} & keV & keV & \rr{\%}\\
      \hline
    O$^{4+}$ & 0.0148 & 0.0004 & \rr{2.7} & 0.0260 & 0.0006 & \rr{2.3} \\
    O$^{5+}$ & 0.0209 & 0.0006 & \rr{2.9} & 0.0331 & 0.0008 & \rr{2.4} \\
    O$^{6+}$ & 0.0271 & 0.0004 & \rr{1.5} & 0.1704 & 0.0052 & \rr{3.1} \\
    O$^{7+}$ & 0.1398 & 0.0052 & \rr{3.7} & 0.2979 & 0.0078 & \rr{2.6} \\
    O$^{8+}$ & 0.2471 & 0.0063 & \rr{2.5} & -- & -- & -- \\
    \hline
    \end{tabular}
    \label{tab:O temp change}
\end{table}

\begin{table}[H]
    \caption{Implied absolute temperature shifts (in keV) \rr{and relative temperature shifts (\%)} from perturbing all Fe ionization and recombination rates with the minimum systematic errors of Tables \ref{tab:ionization} and \ref{tab: recombination} at half peak \rr{ionic concentration} on the unperturbed rising ($T_{rise}$) and falling ($T_{fall}$) CSD curve.}
    \begin{tabular}{ccccccc}
    \hline
      \rr{Ion} & $T_{rise}$ & \multicolumn{2}{c}{\rr{Rising Shift}} & $T_{fall}$ & \multicolumn{2}{c}{\rr{Falling Shift}} \\ 
                & keV & keV & \rr{\%} & keV & keV & \rr{\%}\\
      \hline
    Fe$^{7+}$ & 0.0230 & 0.0009 & \rr{3.9} & 0.0572 & 0.0031 & \rr{5.4} \\
    Fe$^{8+}$ & 0.0402 & 0.0016 & \rr{4.0} & 0.0944 & 0.0036 & \rr{3.8}\\
    Fe$^{9+}$ & 0.0678 & 0.0030 & \rr{4.4} & 0.1238 & 0.0029 & \rr{2.3} \\
    Fe$^{10+}$ & 0.0881 & 0.0030 & \rr{3.4} & 0.1456 & 0.0033 & \rr{2.3} \\
    Fe$^{11+}$ & 0.1065 & 0.0029 & \rr{2.7} & 0.1652 & 0.0026 & \rr{1.6} \\
    Fe$^{12+}$ & 0.1236 & 0.0026 & \rr{2.1} & 0.1850 & 0.0039 & \rr{2.1} \\
    Fe$^{13+}$ & 0.1399 & 0.0026 & \rr{1.9} & 0.2094 & 0.0065 & \rr{3.1} \\
    Fe$^{14+}$ & 0.1563 & 0.0028 & \rr{1.8} & 0.2526 & 0.0180 & \rr{7.1} \\
    Fe$^{15+}$ & 0.1777 & 0.0048 & \rr{2.7} & 0.3935 & 0.0266 & \rr{6.8} \\
    Fe$^{16+}$ & 0.2090 & 0.0065 & \rr{3.1} & 0.6231 & 0.0231 & \rr{3.7} \\
    Fe$^{17+}$ & 0.4092 & 0.0178 & \rr{4.3} & 0.8572 & 0.0344 & \rr{4.0} \\
    Fe$^{18+}$ & 0.5605 & 0.0205 & \rr{3.7} & 1.0023 & 0.0472 & \rr{4.7} \\
    Fe$^{19+}$ & 0.6786 & 0.0340 & \rr{5.0} & 1.1325 & 0.0664 & \rr{5.9} \\
    Fe$^{20+}$ & 0.7820 & 0.0421 & \rr{5.4} & 1.2741 & 0.0749 & \rr{5.9} \\
    Fe$^{21+}$ & 0.8814 & 0.0534 & \rr{6.1} & 1.4660 & 0.0866 & \rr{5.9} \\
    Fe$^{22+}$ & 0.9894 & 0.053 & \rr{5.4} & 1.8313 & 0.1393 & \rr{7.6} \\
    Fe$^{23+}$ & 1.1337 & 0.0478 & \rr{4.2} & 3.2656 & 0.2950 & \rr{9.0} \\
    Fe$^{24+}$ & 1.4706 & 0.0586 & \rr{4.0} & 7.6136 & 0.3788 & \rr{5.0} \\
    Fe$^{25+}$ & 4.7617 & 0.2378 & \rr{5.0} & 23.8883 & 2.7231 & \rr{11.4} \\
    Fe$^{26+}$ & 12.7911 & 0.9198 & \rr{7.2} & -- & -- & -- \\
    \hline
    \end{tabular}
    \label{tab:Fe temp change}
\end{table}

\section{Line Uncertainty Calculation}

\subsection{Method}
Unlike an element's CSD with only $O(Z)$ relevant values, determining ionic line emission requires calculating the ion's energy level population, potentially including thousands or millions of individual calculations. In most cases, no laboratory measurement has ever been made so error estimates rely on comparing different theoretical methods against each other - as done in  \rr{HADP2018} - or extrapolating from the few lab results that do exist. Even if this experimental data were available, theory suggests that some rates will be correlated or anti-correlated \citep{Loch2013TheModels}. Thus creating a table of line emissivities and errors is both impractical and unusable for quantitative analysis.  Nonetheless, observers require some estimate of how uncertainties in atomic data affect models of plasma emission.

For a given ion, the emissivity $\Lambda_{ij}(T_e)$\ of a line with wavelength $\lambda_{ij}$\ due to a transition from level $i \rightarrow j$\ is simply $\Lambda(T_e) = A_{ij} p_i(T_e)$\ where $A_{ij}$\ is the radiative transition rate and $p_i(T_e)$\ the population of level $i$\ at electron temperature $T_e$. Calculating $p_i(T_e)$, however, requires considering all of the processes that populate and depopulate that level; see \citet{Foster2012UpdatedSpectroscopy} and \url{http://www.atomdb.org/physics.php} for details. To probe the non-linear dependence of line emissivities to atomic rate coefficients, we employ \texttt{variableapec} to perturb a single rate coefficient, construct a new collisional radiative matrix, and recalculate line emissivities. We then study how line emissivities vary over a range of perturbation magnitudes and temperatures to understand where atomic uncertainties matter most. \rr{HADP2018} estimated the perturbation amount from differences between atomic codes in collisional excitation rates and transition probabilities. We instead estimate a perturbation amount associated with the uncertainty from atomic calculations of rate coefficients or experimental errors. Note that we quantify the effect of atomic uncertainties on line emissivities by calculating changes to the line emissivity. We use the word ``change'' rather than ``error''  to reference line emissivities (or other values) we have calculated theoretically, as we employ the word ``error'' in reference to measured errors (\textit{e.g.} errors on CSD rates or experimental errors on transition rates). Uncertainties are used throughout to mean general unknowns in the fundamental atomic data.  

\rr{The \texttt{variableapec} code will perturb rate coefficients for collisional excitation to an individual level ($i$)\ or radiative transition rates $A_{ij}$\ at the temperature where the ion's concentration peaks. It will then determine the fractional changes in emissivity for all transitions from level $i$\ as well as any others affected. The code outputs the fractional change in emissivity d$\epsilon$/$\epsilon$, \textit{i.e.} the averaged fractional change in emissivity calculated from the positive and negative error on the rate. It also calculates the partial derivatives of the resulting line emissivities ($\epsilon$) with respect to the change in the rate $d\epsilon/dr$. This value shows how sensitive each line is to uncertainties in the underlying rates.}

For this phase of the study, we considered the low density case only. At densities where the electronic collision cross section competes with the radiative transition rate, it is necessary to also consider perturbations to the $A$ value. We found that at these larger densities such perturbations can affect thousands of lines by an observable amount and thus we leave this project for a future paper.

\subsection{Results}

The challenge of calculating line emissivities in a collisional plasma varies substantially depending upon the line. An electric dipole transition involving low-lying excited levels, such as the He-like resonance line $1s2p{}^1$P$_1 \rightarrow 1s^2 {}^1$S$_0$, may have an emission entirely dominated by direct collisional excitation from the ground state followed by a radiative transition to ground. As typical $A$ values for such transitions are large, at densities below $10^8$\,cm$^{-3}$, the line emissivity uncertainty is governed by this one excitation cross section. More complex situations, however, are common as an excited level may be populated by excitation ionization, recombination, and cascades as well as direct excitation. Determining experimental uncertainties for each of these rates is impractical, and in most cases even theoretical uncertainties are estimated simply by comparing the values obtained from using different methods. 

We therefore focus on understanding the variation in the final emissivities as the underlying excitation or radiative rates are changed, expressed as partial derivatives. The {\tt variableapec}\ code can calculate these for any transition in the AtomDB; we provide examples here for select transitions in the \ion{Fe}{xvii} and \ion{Fe}{xix} line complexes. Despite the fact these key transitions in particular have been poorly measured historically, \ion{Fe}{xvii} and \ion{Fe}{xix} have strong lines that are commonly used temperature diagnostics \citep{Gu2020X-rayCapella}. We perturb the atomic data driving strong lines in \ion{Fe}{xvii} and \ion{Fe}{xix} by 5\% and 20\% and compare fractional emissivity changes. At low density, most line emissivities are not sensitive to perturbations in $A$ value. For example, while perturbations to the $A$ value of strong \ion{Fe}{xvii} lines do not affect any other lines, varying the $A$ value of two \ion{Fe}{xix} lines by 20\% affects those lines by only 2\% but two other X-ray lines by 20\%. However, a 20\% perturbation to the direct excitation rate of a strong line at \rr{the temperature of peak ionic concentration} can increase both that line's emissivity and others up to 20\%.  

We varied the direct excitation rate (from ground) for a number of \ion{Fe}{xvii} lines (3C, 3D, 3E, 3F, M2, 3G) and found that, out of this group of lines, only the line whose rate was perturbed changed. Varying the rates by 5\% changed the M2, 3G, and 3F lines by 1\% or less and the 3E, 3D, and 3C lines by 4-5\%, and proportionally larger for 20\% changes. For the 3E, 3D, and 3C lines, a 20\% perturbation to the direct excitation rates also led to 16-20\% variations in some strong EUV (100-500\AA), although no X-ray lines were significantly affected. 

We then perturbed the direct excitation rates (again from ground) by 20\% for four strong lines of \ion{Fe}{xix}: O19 (53$\rightarrow$1), O24 (68$\rightarrow$1), O25 (71$\rightarrow$1), O26 (74$\rightarrow$1), using the nomenclature of \citet{Brown2002LaboratoryEmission} with the AtomDB v3.0.9 levels in parentheses. As with \ion{Fe}{xvii}, the emissivity of these strong lines only changes when their own direct excitation rate is varied and by 17-20\%.

\begin{figure}[H]
\includegraphics[width=8.5 cm]{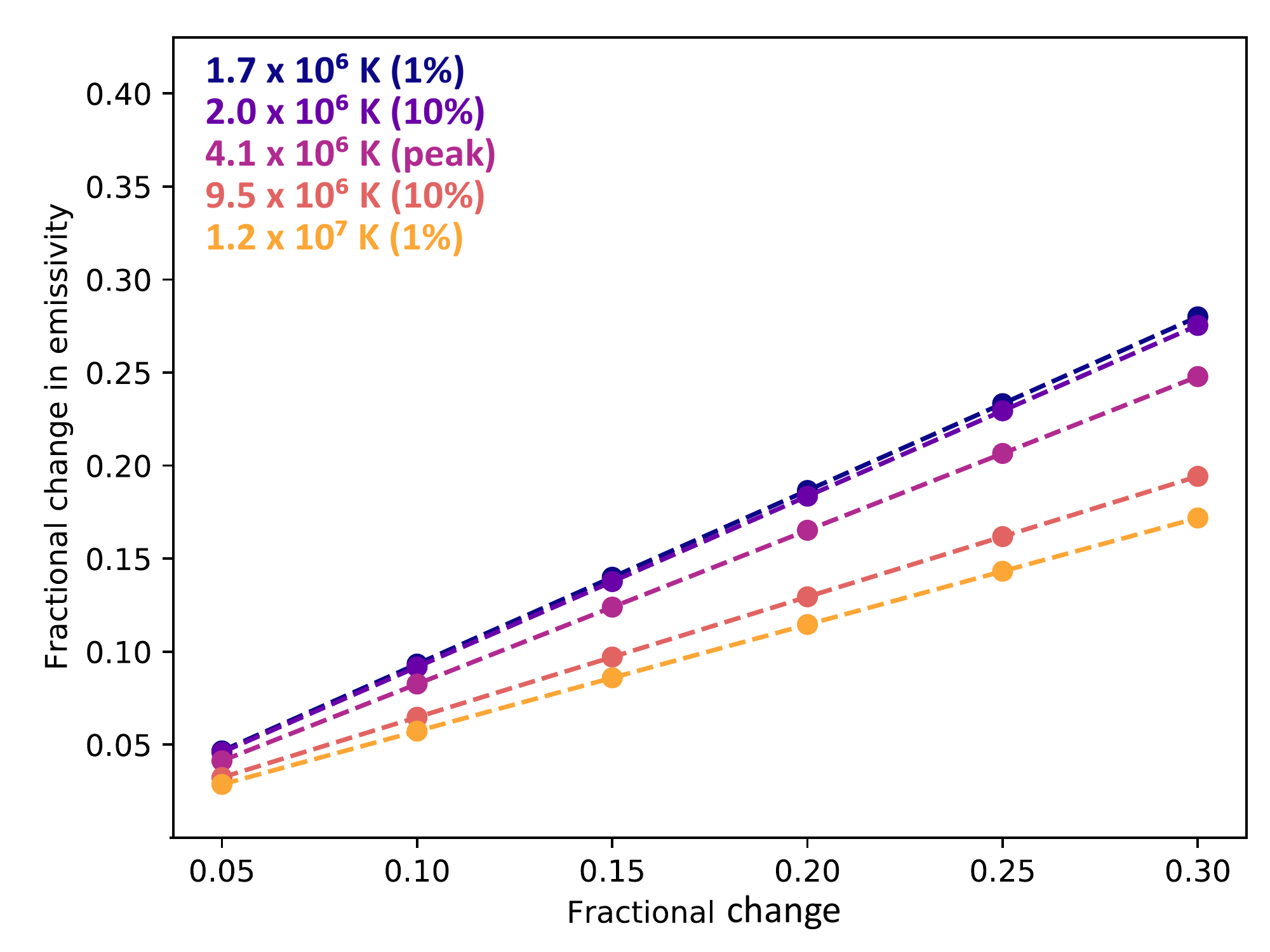}
\includegraphics[width=8.5 cm]{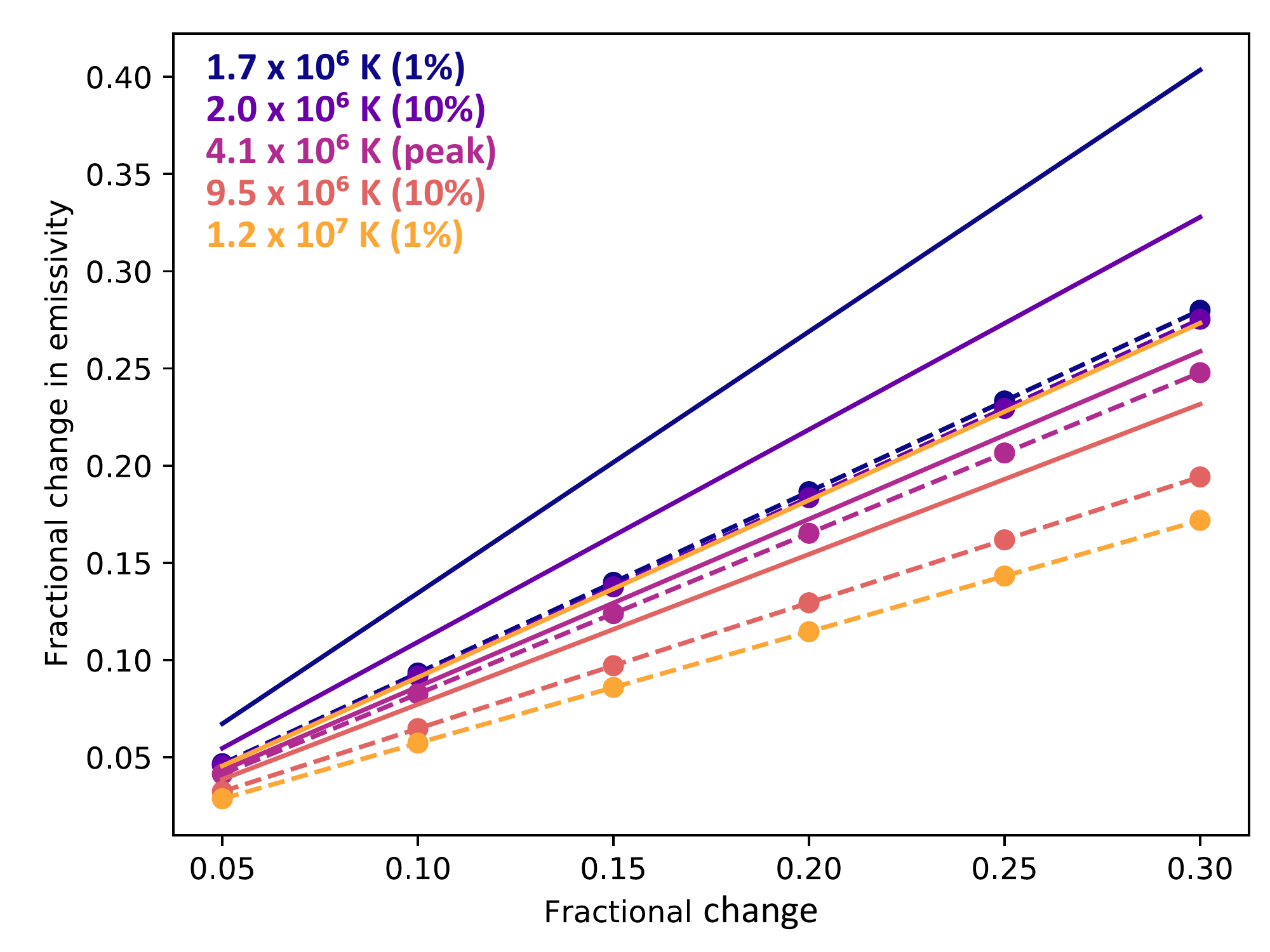}
\caption{[Left] \ion{Fe}{xvii} 17$\rightarrow$6 sensitivity over various temperatures at low density due only to changes to the 17$\rightarrow$1 direct excitation rate. [Right] Same (dashed) compared to the sensitivity from also including the maximum perturbed CSD due to systematics (solid).}
\label{fig:Fe_line_sens}
\end{figure}

The partial derivative analysis reveals additional lines affected by a perturbation to the direct excitation rate of a single line. To investigate the correlation of line emissivity errors, we study the effect of various magnitudes of errors on emissivity changes over a range of temperatures. For \ion{Fe}{xvii}, we found that the EUV line at 263.63 \AA\ is the strongest line impacted by perturbations to the direct excitation rate of the 3E line (17$\rightarrow$1) (Figure \ref{fig:Fe_line_sens} [Left]). At peak ion temperature, varying the direct excitation rate of the 3E line by 20\% results in a 17\% change in the emissivity (1.79$\times10^{-13}$ ph cm$^3$ s$^{-1}$) of the 17$\rightarrow$6 transition (263.63 \AA). We perturb the 3E line by various magnitudes at the designated five temperatures of interest (see Figure~\ref{fig:5_temps}) to study the correlation between these two lines. The fractional change in emissivity is greatest at low temperatures where the \rr{ionic concentration} is between 1-10\%. The maximum difference in fractional change in emissivity across the temperature range increases with perturbation magnitude. We see that the 17$\rightarrow$6 line emissivity is then sensitive to both temperature and any uncertainties in the 3E direct excitation rate.

We repeat the partial derivative analysis for \ion{Fe}{xix} to look for any X-ray lines that may be affected by perturbations to the direct excitation rates of key \ion{Fe}{xix} lines. We find that the O20 line from \cite{Brown2002LaboratoryEmission} (74$\rightarrow$4) is affected by an observable amount with perturbations to the direct excitation rate of O25 (74$\rightarrow$1). Unlike the \ion{Fe}{xvii} line at 263.63 \AA, this line is not temperature sensitive: the fractional change in the emissivity of O20 is approximately equal to the error on the direct excitation rate of O25 for all temperatures. This is a subtle but important detail in that it shows that the O20 line is primarily populated by direct excitation from the ground level. If recombination or ionization contributed more to the line emissivity, the temperature of the plasma would be more significant. Conversely, ionization and recombination of \ion{Fe}{xvi} and \ion{Fe}{xviii} into \ion{Fe}{xvii} contributes to the 17$\rightarrow$6 line emissivity, so the plasma temperature factors significantly into the fractional change in emissivity. 

This is evident in Figure \ref{fig:Fe_line_sens} [Right], where we show the \ion{Fe}{xvii} 17$\rightarrow$6 line sensitivity to the 17$\rightarrow$1 direct excitation rate as well as the sensitivity from employing the maximum possible CSD values due to the typical systematics listed in Tables \ref{tab:ionization} and \ref{tab: recombination} in emissivity calculations. We find that the CSD error, \textit{i.e.} the contribution of the perturbed CSD to the increase in fractional change in emissivity from the non-perturbed case, is greatest at temperatures where the \rr{ionic concentration} is between 1-10\% and smallest at the temperature of peak \rr{ionic concentration}. This matches expectations given the sensitivity \rr{of ionic concentrations} to perturbations to the CSD shown in Figure \ref{fig:new_csd}.

\begin{figure}
\centering
\includegraphics[width=8.5 cm]{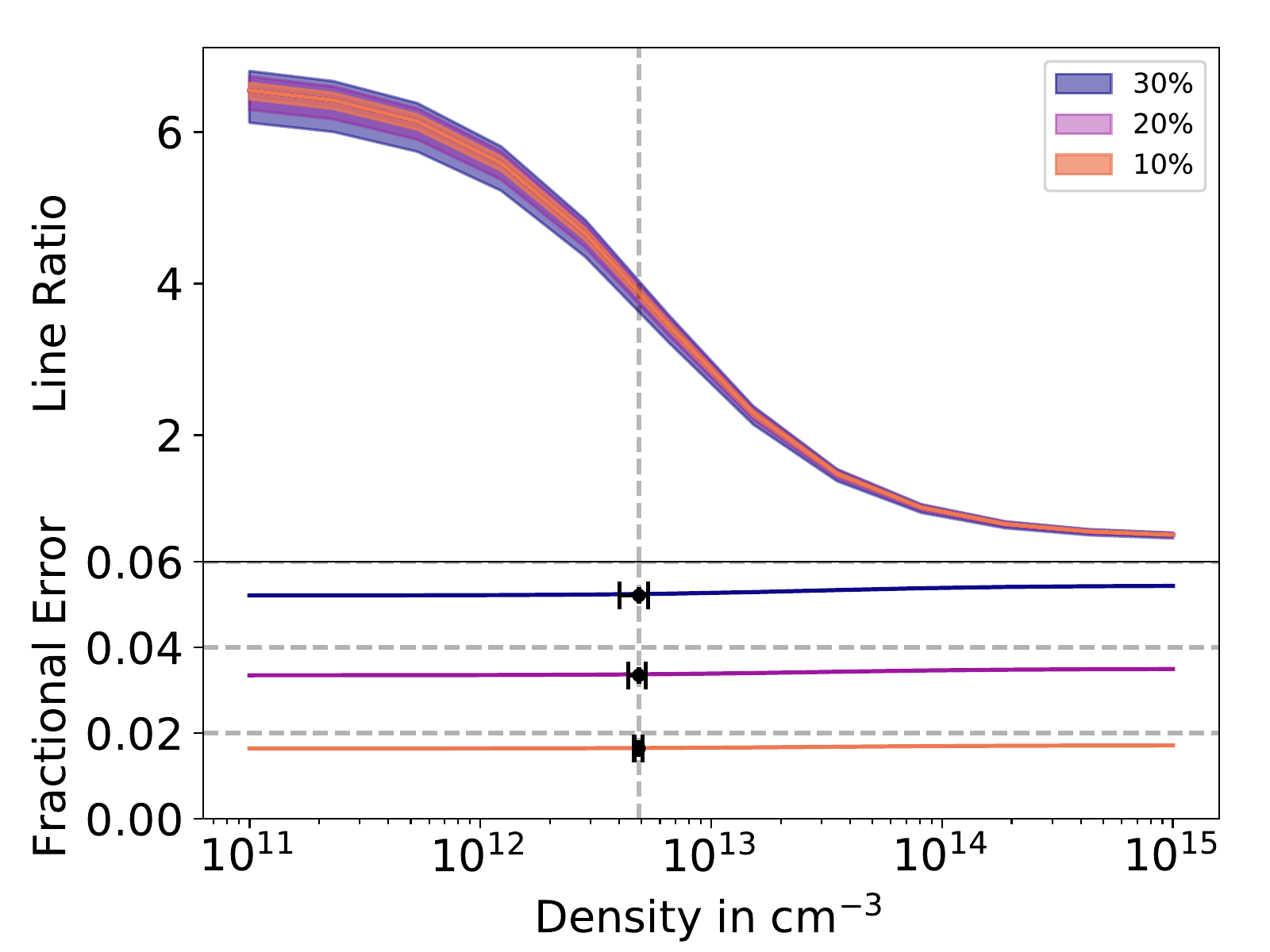}
\caption{The \ion{Fe}{xxii} {\it I}(11.92 \AA)/{\it I}(11.77 \AA) line ratio at $T_e = 12.8\times10^{6}$ K for three different changes to the $A$\ value of the transitions. \rr{The error bars on the bottom panel show the uncertainty of the density diagnostic for when the line ratio equals 4.}}
\label{fig:Fe_22}
\end{figure}

We also investigate the effect of atomic uncertainties on density diagnostics for line ratios, such as the density-sensitive \ion{Fe}{xxii} {\it I}(11.92 \AA)/{\it I}(11.77 \AA) line ratio as studied in \citet{2003ApJ...588L.101M} and \citet{2006Natur.441..953M}. We changed the $A$ values of the two lines individually by 10\%, 20\% and 30\% and recalculated the final line ratio, finding a resulting change to the final line ratio that is about a sixth of the perturbation magnitude in all cases \rr{[see Figure \ref{fig:Fe_22}]}. Unlike other temperature-sensitive line ratios, the final fractional change is not density-sensitive. \rr{Perturbing these two lines, however, can result in a relative $\pm$ error on density diagnostics that is asymmetric. Looking at the density where the line ratio equals 4, a 10\% perturbation gives a (-4.6\%, +3.8\%) error on the unperturbed density of $4.86\times10^{13}$ cm$^{-3}$, while 20\% and 30\% perturbations give (-10.3\%, +6.9\%) and (-17.7\%, +9.6\%) errors respectively.}

\section{Discussion and Conclusions} 
The problem of providing EUV/X-ray spectroscopists with practical error estimates has been long-standing and will only become more acute as newer facilities come online with higher resolution and more sensitivity. Unfortunately, no theoretical method yet exists that can generate reliable error estimates for atomic calculations, including the effect of correlations. The goal of this paper is not to solve the problem but rather to provide observers with practical methods to estimate the uncertainty involved in a specific calculation. We also hope to motivate further and more accurate lab measurements of atomic rates, a problem that requires significant and long-term funding. As our results show, it is important to consider all the parameters that contribute to an analysis, not merely statistical errors on the spectrum. 

Our primary focus was uncertainties in CSD calculations as this involves a limited number of rates, most of which have at least some experimental measurements. In conducting a literature search for rate coefficients and systematic uncertainties, we found that while a number of papers explicitly detail the components of their systematic errors and the justification for each value, more than a few provide little to no such information. We admit to being surprised, however, that even using only the published systematic errors showed that most modern CSD calculations have converged to lie within their bounds. It will be difficult to make significant improvements, then, without more accurate measurements for theory to attempt to match. Of course, this does not test the effect of processes that have been so far excluded from the CSD calculation entirely, such as electron-impact multiple-ionization effects \citep{2017ApJ...850..122H}. Direct measurements of CSDs as a function of temperature would be most useful to ensure the CSD calculations contain all the key processes.

We have also investigated the uncertainties involved in the calculation of emission line fluxes. The {\tt variableapec}\ code allows users to recreate calculations with a range of perturbations in order to find the lines most sensitive to atomic uncertainties and which uncertainties are correlated. The code will also show the different components that enter into the line flux beyond simple direct excitation, such as ionization, recombination, and cascade effects. While we only focused here on lines sensitive to perturbations in excitation rates, emission and absorption lines will also be affected by changes to $A$ values, especially when using lines as density diagnostics. 
 
Our methods aim to address observers' needs by explicitly showing the effect individual uncertainties have on a final error. In some cases, it can also show how specific line ratios are less affected by correlations or errors than others, aiding in analysis even if direct laboratory measurements are not available. This will also motivate identification of specific atomic data that requires more precise lab measurements or theoretical atomic calculations for better accuracy. Ultimately, optimizing spectral modeling to extract the maximum from an observed spectrum will involve determining which atomic data are ill-measured, the ions most sensitive to atomic uncertainties, and the temperature regions where uncertainties most greatly impact observable plasma features.

\rr{We hope the \texttt{variableapec} tools developed for this paper will be useful for people to calculate their own estimate uncertainties given their own interpretations of underlying experimental uncertainties [see Appendix for examples of routine commands]. We have completed the CSD error estimates for Fe and O and made these available as part of the Github distribution. We invite contributions to the open-source project to add other ions and elements to the database.}

\acknowledgements
The authors gratefully acknowledge support from NASA Astrophysics Grant 80NSSC18K0409.

\bibliographystyle{aasjournal}
\bibliography{main}

\begin{thebibliography}{}
\expandafter\ifx\csname natexlab\endcsname\relax\def\natexlab#1{#1}\fi
\providecommand{\url}[1]{\href{#1}{#1}}
\providecommand{\dodoi}[1]{doi:~\href{http://doi.org/#1}{\nolinkurl{#1}}}
\providecommand{\doeprint}[1]{\href{http://ascl.net/#1}{\nolinkurl{http://ascl.net/#1}}}
\providecommand{\doarXiv}[1]{\href{https://arxiv.org/abs/#1}{\nolinkurl{https://arxiv.org/abs/#1}}}

\bibitem[{Aichele {et~al.}(1998)Aichele, Hartenfeller, Hathiramani, Hofmann,
  Sch{\"{a}}fer, Steidl, Stenke, Salzborn, Pattard, \&
  Rost}]{Aichele1998ElectronO7+}
Aichele, K., Hartenfeller, U., Hathiramani, D., {et~al.} 1998, Journal of
  Physics B Atomic Molecular Physics, 31, 2369,
  \dodoi{10.1088/0953-4075/31/10/023}

\bibitem[{{Altun} {et~al.}(2007){Altun}, {Yumak}, {Yavuz}, {Badnell}, {Loch},
  \& {Pindzola}}]{Altun2007DRMagnesium}
{Altun}, Z., {Yumak}, A., {Yavuz}, I., {et~al.} 2007, \aap, 474, 1051,
  \dodoi{10.1051/0004-6361:20078238}

\bibitem[{{Andersen} \& {Bolko}(1990)}]{1990JPhB...23.3167A}
{Andersen}, L.~H., \& {Bolko}, J. 1990, Journal of Physics B Atomic Molecular
  Physics, 23, 3167, \dodoi{10.1088/0953-4075/23/18/019}

\bibitem[{{Andersen} {et~al.}(1990){Andersen}, {Bolko}, \&
  {Kvistgaard}}]{1990PhRvA..41.1293A}
{Andersen}, L.~H., {Bolko}, J., \& {Kvistgaard}, P. 1990, PRA, 41, 1293,
  \dodoi{10.1103/PhysRevA.41.1293}

\bibitem[{Arnaud \& Raymond(1992)}]{Arnaud1992IronEquilibrium}
Arnaud, M., \& Raymond, J. 1992, ApJ, 398, 394, \dodoi{10.1086/171864}

\bibitem[{Arnaud \& Rothenflug(1985)}]{Arnaud1985AnRates}
Arnaud, M., \& Rothenflug, R. 1985, A{\&}AS, 60, 425

\bibitem[{Badnell(2006)}]{Badnell2006RadiativePlasmas}
Badnell, N.~R. 2006, ApJS, 167, 334, \dodoi{10.1086/508465}

\bibitem[{Bell \& Bell(1982)}]{Bell:1981vja}
Bell, M., \& Bell, J. 1982, Part. Accel., 12, 49

\bibitem[{Bernhardt {et~al.}(2014)Bernhardt, Becker, Grieser, Hahn, Krantz,
  Lestinsky, Novotn{\'{y}}, Repnow, Savin, Spruck, Wolf, M{\"{u}}ller, \&
  Schippers}]{Bernhardt2014AbsoluteRing}
Bernhardt, D., Becker, A., Grieser, M., {et~al.} 2014, Phys. Rev. A, 90, 12702,
  \dodoi{10.1103/PhysRevA.90.012702}

\bibitem[{B{\"{o}}hm {et~al.}(2002)B{\"{o}}hm, Schippers, Shi, M{\"{u}}ller,
  Ekl{\"{o}}w, Schuch, Danared, Badnell, Mitnik, \&
  Griffin}]{Bohm2002MeasurementN}
B{\"{o}}hm, S., Schippers, S., Shi, W., {et~al.} 2002, Physical Review A, 65,
  52728, \dodoi{10.1103/PhysRevA.65.052728}

\bibitem[{Brooks {et~al.}(1980)Brooks, Datla, Krumbein, \&
  Griem}]{Brooks1980MeasurementXi}
Brooks, R.~L., Datla, R.~U., Krumbein, A.~D., \& Griem, H.~R. 1980, Physical
  Review A, 21, 1387, \dodoi{10.1103/PhysRevA.21.1387}

\bibitem[{Brown {et~al.}(2002)Brown, Beiersdorfer, Liedahl, Widmann, Kahn, \&
  Clothiaux}]{Brown2002LaboratoryEmission}
Brown, G.~V., Beiersdorfer, P., Liedahl, D.~A., {et~al.} 2002, ApJS, 140, 589,
  \dodoi{10.1086/339374}

\bibitem[{Bryans {et~al.}(2006)Bryans, Badnell, Gorczyca, Laming, Mitthumsiri,
  \& Savin}]{Bryans2006CollisionalIons}
Bryans, P., Badnell, N.~R., Gorczyca, T.~W., {et~al.} 2006, ApJ, 167, 343,
  \dodoi{10.1086/507629}

\bibitem[{Bryans {et~al.}(2009)Bryans, Landi, \&
  Savin}]{Bryans2009ACoefficients}
Bryans, P., Landi, E., \& Savin, D.~W. 2009, ApJ, 691, 1540,
  \dodoi{10.1088/0004-637X/691/2/1540}

\bibitem[{Chen(1988)}]{Chen1988DielectronicCalculations}
Chen, M.~H. 1988, Phys Rev A, 38, 2332, \dodoi{10.1103/PhysRevA.38.2332}

\bibitem[{Chen {et~al.}(1990)Chen, Reed, \&
  Moores}]{Chen1990ContributionsFe15+}
Chen, M.~H., Reed, K.~J., \& Moores, D.~L. 1990, Phys. Rev. Letters, 64, 1350,
  \dodoi{10.1103/PhysRevLett.64.1350}

\bibitem[{Crandall {et~al.}(1986)Crandall, Phaneuf, Gregory, Howald, \&
  Mueller}]{Crandall1986Electron-impactIons}
Crandall, D.~H., Phaneuf, R.~A., Gregory, D.~C., Howald, A.~M., \& Mueller,
  D.~W. 1986, Phys. Rev. A, 34, 1757, \dodoi{10.1103/PhysRevA.34.1757}

\bibitem[{Dere(2007)}]{Dere2007IonizationZinc}
Dere, K.~P. 2007, A{\&}A, 466, 771, \dodoi{10.1051/0004-6361:20066728}

\bibitem[{Dittner {et~al.}(1987{\natexlab{a}})Dittner, Datz, Miller, Pepmiller,
  \& Fou}]{Dittner1987DielectronicO5+}
Dittner, P.~F., Datz, S., Miller, P.~D., Pepmiller, P.~L., \& Fou, C.~M.
  1987{\natexlab{a}}, Phys. Rev. A, 35, 3668, \dodoi{10.1103/PhysRevA.35.3668}

\bibitem[{Dittner {et~al.}(1987{\natexlab{b}})Dittner, Datz, Krause, Miller,
  Pepmiller, Bottcher, Fou, Griffin, \& Pindzola}]{Dittner1987DielectronicF5+}
Dittner, P.~F., Datz, S., Krause, H.~F., {et~al.} 1987{\natexlab{b}}, Physical
  Review A, 36, 33, \dodoi{10.1103/PhysRevA.36.33}

\bibitem[{Falk {et~al.}(1983)Falk, Stefani, Camilloni, Dunn, Phaneuf, Gregory,
  \& Crandall}]{Falk1983Measured+}
Falk, R.~A., Stefani, G., Camilloni, R., {et~al.} 1983, Phys. Rev. A, 28, 91,
  \dodoi{10.1103/PhysRevA.28.91}

\bibitem[{Fogle {et~al.}(2008)Fogle, Bahati, Bannister, Vane, Loch, Pindzola,
  Ballance, Thomas, Zhaunerchyk, Bryans, Mitthumsiri, \&
  Savin}]{Fogle2008Electron-ImpactV}
Fogle, M., Bahati, E.~M., Bannister, M.~E., {et~al.} 2008, ApJS, 175, 543,
  \dodoi{10.1086/525256}

\bibitem[{Foster {et~al.}(2012)Foster, Ji, Smith, \&
  Brickhouse}]{Foster2012UpdatedSpectroscopy}
Foster, A.~R., Ji, L., Smith, R.~K., \& Brickhouse, N.~S. 2012, ApJ, 756, 128,
  \dodoi{10.1088/0004-637X/756/2/128}

\bibitem[{{Golden} \& {Sampson}(1977)}]{1977JPhB...10.2229G}
{Golden}, L.~B., \& {Sampson}, D.~H. 1977, Journal of Physics B Atomic
  Molecular Physics, 10, 2229, \dodoi{10.1088/0022-3700/10/11/023}

\bibitem[{Gregory {et~al.}(1987)Gregory, Wang, Meyer, \&
  Rinn}]{Gregory1987Electron-impact+}
Gregory, D.~C., Wang, L.~J., Meyer, F.~W., \& Rinn, K. 1987, Phys. Rev. A, 35,
  3256, \dodoi{10.1103/PhysRevA.35.3256}

\bibitem[{Gu {et~al.}(2020)Gu, Shah, Mao, Raassen, de~Plaa, Pinto, Akamatsu,
  Werner, Simionescu, Mernier, Sawada, Mohanty, Amaro, Gu, Porter, Crespo
  L{\'{o}}pez-Urrutia, \& Kaastra}]{Gu2020X-rayCapella}
Gu, L., Shah, C., Mao, J., {et~al.} 2020, arXiv e-prints, arXiv:2007.03843

\bibitem[{Gwinner {et~al.}(2001)Gwinner, Savin, Schwalm, Wolf, Schippers,
  M{\"{u}}ller, Badnell, \& Chen}]{Gwinner2001DielectronicIons}
Gwinner, G., Savin, D.~W., Schwalm, D., {et~al.} 2001, Physica Scripta Volume
  T, 92, 319, \dodoi{10.1238/Physica.Topical.092a00319}

\bibitem[{Hahn(2014)}]{Hahn2014ElectronIons}
Hahn, M. 2014, in Journal of Physics Conference Series, Vol. 488, Journal of
  Physics Conference Series, 12050, \dodoi{10.1088/1742-6596/488/1/012050}

\bibitem[{{Hahn} {et~al.}(2017){Hahn}, {M{\"u}ller}, \&
  {Savin}}]{2017ApJ...850..122H}
{Hahn}, M., {M{\"u}ller}, A., \& {Savin}, D.~W. 2017, \apj, 850, 122,
  \dodoi{10.3847/1538-4357/aa9276}

\bibitem[{Hahn {et~al.}(2011{\natexlab{a}})Hahn, Bernhardt, Grieser, Krantz,
  Lestinsky, M{\"{u}}ller, Novotn{\'{y}}, Repnow, Schippers, Wolf, \&
  Savin}]{Hahn2011StorageFe13+}
Hahn, M., Bernhardt, D., Grieser, M., {et~al.} 2011{\natexlab{a}}, ApJ, 729,
  76, \dodoi{10.1088/0004-637X/729/1/76}

\bibitem[{Hahn {et~al.}(2011{\natexlab{b}})Hahn, Grieser, Krantz, Lestinsky,
  M{\"{u}}ller, Novotn{\'{y}}, Repnow, Schippers, Wolf, \&
  Savin}]{Hahn2011Storage+}
Hahn, M., Grieser, M., Krantz, C., {et~al.} 2011{\natexlab{b}}, ApJ, 735, 105,
  \dodoi{10.1088/0004-637X/735/2/105}

\bibitem[{{Hahn} {et~al.}(2012){Hahn}, {Becker}, {Grieser}, {Krantz},
  {Lestinsky}, {M{\"u}ller}, {Novotn{\'y}}, {Repnow}, {Schippers}, {Spruck},
  {Wolf}, \& {Savin}}]{2012ApJ...760...80H}
{Hahn}, M., {Becker}, A., {Grieser}, M., {et~al.} 2012, \apj, 760, 80,
  \dodoi{10.1088/0004-637X/760/1/80}

\bibitem[{Hahn {et~al.}(2013)Hahn, Becker, Bernhardt, Grieser, Krantz,
  Lestinsky, M{\"{u}}ller, Novotn{\'{y}}, Repnow, Schippers, Spruck, Wolf, \&
  Savin}]{Hahn2013StorageAstrophysics}
Hahn, M., Becker, A., Bernhardt, D., {et~al.} 2013, in American Astronomical
  Society Meeting Abstracts, Vol. 222, American Astronomical Society Meeting
  Abstracts, 114.06

\bibitem[{{Hahn} {et~al.}(2015){Hahn}, {Becker}, {Bernhardt}, {Grieser},
  {Krantz}, {Lestinsky}, {M{\"u}ller}, {Novotn{\'y}}, {Repnow}, {Schippers},
  {Spruck}, {Wolf}, \& {Savin}}]{2015ApJ...813...16H}
{Hahn}, M., {Becker}, A., {Bernhardt}, D., {et~al.} 2015, \apj, 813, 16,
  \dodoi{10.1088/0004-637X/813/1/16}

\bibitem[{Hahn {et~al.}(2016)Hahn, Becker, Bernhardt, Grieser, Krantz,
  Lestinsky, M{\"{u}}ller, Novotn{\'{y}}, Pindzola, Repnow, Schippers, Spruck,
  Wolf, \& Savin}]{Hahn2016StorageFe8+}
Hahn, M., Becker, A., Bernhardt, D., {et~al.} 2016, Journal of Physics B Atomic
  Molecular Physics, 49, 84006, \dodoi{10.1088/0953-4075/49/8/084006}

\bibitem[{{Hitomi Collaboration} {et~al.}(2016){Hitomi Collaboration},
  Aharonian, Akamatsu, Akimoto, Allen, Anabuki, Angelini, Arnaud, Audard,
  Awaki, Axelsson, Bamba, Bautz, Blandford, Brenneman, Brown, Bulbul, Cackett,
  Chernyakova, Chiao, Coppi, Costantini, de~Plaa, den Herder, Done, Dotani,
  Ebisawa, Eckart, Enoto, Ezoe, Fabian, Ferrigno, Foster, Fujimoto, Fukazawa,
  Furuzawa, Galeazzi, Gallo, Gandhi, Giustini, Goldwurm, Gu, Guainazzi, Haba,
  Hagino, Hamaguchi, Harrus, Hatsukade, Hayashi, Hayashi, Hayashida, Hiraga,
  Hornschemeier, Hoshino, Hughes, Iizuka, Inoue, Inoue, Ishibashi, Ishida,
  Ishikawa, Ishisaki, Itoh, Iyomoto, Kaastra, Kallman, Kamae, Kara, Kataoka,
  Katsuda, Katsuta, Kawaharada, Kawai, Kelley, Khangulyan, Kilbourne, King,
  Kitaguchi, Kitamoto, Kitayama, Kohmura, Kokubun, Koyama, Koyama, Kretschmar,
  Krimm, Kubota, Kunieda, Laurent, Lebrun, Lee, Leutenegger, Limousin,
  Loewenstein, Long, Lumb, Madejski, Maeda, Maier, Makishima, Markevitch,
  Matsumoto, Matsushita, McCammon, McNamara, Mehdipour, Miller, Miller,
  Mineshige, Mitsuda, Mitsuishi, Miyazawa, Mizuno, Mori, Mori, Moseley, Mukai,
  Murakami, Murakami, Mushotzky, Nagino, Nakagawa, Nakajima, Nakamori, Nakano,
  Nakashima, Nakazawa, Nobukawa, Noda, Nomachi, O'Dell, Odaka, Ohashi, Ohno,
  Okajima, Ota, Ozaki, Paerels, Paltani, Parmar, Petre, Pinto, Pohl, Porter,
  Pottschmidt, Ramsey, Reynolds, Russell, Safi-Harb, Saito, Sakai, Sameshima,
  Sato, Sato, Sato, Sawada, Schartel, Serlemitsos, Seta, Shidatsu, Simionescu,
  Smith, Soong, Stawarz, Sugawara, Sugita, Szymkowiak, Tajima, Takahashi,
  Takahashi, Takeda, Takei, Tamagawa, Tamura, Tamura, Tanaka, Tanaka, Tanaka,
  Tashiro, Tawara, Terada, Terashima, Tombesi, Tomida, Tsuboi, Tsujimoto,
  Tsunemi, Tsuru, Uchida, Uchiyama, Uchiyama, Ueda, Ueda, Ueno, Uno, Urry,
  Ursino, de~Vries, Watanabe, Werner, Wik, Wilkins, Williams, Yamada,
  Yamaguchi, Yamaoka, Yamasaki, Yamauchi, Yamauchi, Yaqoob, Yatsu, Yonetoku,
  Yoshida, Yuasa, Zhuravleva, \& Zoghbi}]{HitomiCollaboration2016TheCluster}
{Hitomi Collaboration}, Aharonian, F., Akamatsu, H., {et~al.} 2016, Nature,
  535, 117, \dodoi{10.1038/nature18627}

\bibitem[{{Hitomi Collaboration} {et~al.}(2018){Hitomi Collaboration},
  Aharonian, Akamatsu, Akimoto, Allen, Angelini, Audard, Awaki, Axelsson,
  Bamba, Bautz, Blandford, Brenneman, Brown, Bulbul, Cackett, Chernyakova,
  Chiao, Coppi, Costantini, de~Plaa, de~Vries, den Herder, Done, Dotani,
  Ebisawa, Eckart, Enoto, Ezoe, Fabian, Ferrigno, Foster, Fujimoto, Fukazawa,
  Furuzawa, Galeazzi, Gallo, Gandhi, Giustini, Goldwurm, Gu, Guainazzi, Haba,
  Hagino, Hamaguchi, Harrus, Hatsukade, Hayashi, Hayashi, Hayashida, Hell,
  Hiraga, Hornschemeier, Hoshino, Hughes, Ichinohe, Iizuka, Inoue, Inoue,
  Ishida, Ishikawa, Ishisaki, Iwai, Kaastra, Kallman, Kamae, Kataoka, Katsuda,
  Kawai, Kelley, Kilbourne, Kitaguchi, Kitamoto, Kitayama, Kohmura, Kokubun,
  Koyama, Koyama, Kretschmar, Krimm, Kubota, Kunieda, Laurent, Lee,
  Leutenegger, Limousin, Loewenstein, Long, Lumb, Madejski, Maeda, Maier,
  Makishima, Markevitch, Matsumoto, Matsushita, McCammon, McNamara, Mehdipour,
  Miller, Miller, Mineshige, Mitsuda, Mitsuishi, Miyazawa, Mizuno, Mori, Mori,
  Mukai, Murakami, Mushotzky, Nakagawa, Nakajima, Nakamori, Nakashima,
  Nakazawa, Nobukawa, Nobukawa, Noda, Odaka, Ohashi, Ohno, Okajima, Ota, Ozaki,
  Paerels, Paltani, Petre, Pinto, Porter, Pottschmidt, Reynolds, Safi-Harb,
  Saito, Sakai, Sasaki, Sato, Sato, Sato, Sawada, Schartel, Serlemtsos, Seta,
  Shidatsu, Simionescu, Smith, Soong, Stawarz, Sugawara, Sugita, Szymkowiak,
  Tajima, Takahashi, Takahashi, Takeda, Takei, Tamagawa, Tamura, Tanaka,
  Tanaka, Tanaka, Tashiro, Tawara, Terada, Terashima, Tombesi, Tomida, Tsuboi,
  Tsujimoto, Tsunemi, Tsuru, Uchida, Uchiyama, Uchiyama, Ueda, Ueda, Uno, Urry,
  Ursino, Watanabe, Werner, Wilkins, Williams, Yamada, Yamaguchi, Yamaoka,
  Yamasaki, Yamauchi, Yamauchi, Yaqoob, Yatsu, Yonetoku, Zhuravleva, Zoghbi, \&
  Raassen}]{HitomiCollaboration2018AtomicHitomi}
---. 2018, PASJ, 70, 12, \dodoi{10.1093/pasj/psx156}

\bibitem[{Holt {et~al.}(1979)Holt, White, Becker, Boldt, Mushotzky,
  Serlemitsos, \& Smith}]{Holt1979X-rayCapella.}
Holt, S.~S., White, N.~E., Becker, R.~H., {et~al.} 1979, ApJL, 234, L65,
  \dodoi{10.1086/183110}

\bibitem[{Huenemoerder {et~al.}(2003)Huenemoerder, Canizares, Drake, \&
  Sanz-Forcada}]{Huenemoerder2003TheLacertae}
Huenemoerder, D.~P., Canizares, C.~R., Drake, J.~J., \& Sanz-Forcada, J. 2003,
  ApJ, 595, 1131, \dodoi{10.1086/377490}

\bibitem[{{Jacobs} {et~al.}(1977){Jacobs}, {Davis}, {Kepple}, \&
  {Blaha}}]{1977ApJ...211..605J}
{Jacobs}, V.~L., {Davis}, J., {Kepple}, P.~C., \& {Blaha}, M. 1977, \apj, 211,
  605, \dodoi{10.1086/154970}

\bibitem[{Jakubowicz \& Moores(1981)}]{Jakubowicz1981ElectronIons}
Jakubowicz, H., \& Moores, D.~L. 1981, Journal of Physics B Atomic Molecular
  Physics, 14, 3733, \dodoi{10.1088/0022-3700/14/19/022}

\bibitem[{Jordan(1969)}]{Jordan1969TheNickel}
Jordan, C. 1969, MNRAS, 142, 501, \dodoi{10.1093/mnras/142.4.501}

\bibitem[{Kaastra {et~al.}(1996)Kaastra, Mewe, \&
  Nieuwenhuijzen}]{Kaastra1996SPEX:Spectra.}
Kaastra, J.~S., Mewe, R., \& Nieuwenhuijzen, H. 1996, in UV and X-ray
  Spectroscopy of Astrophysical and Laboratory Plasmas, 411--414

\bibitem[{Kallne \& Jones(1977)}]{Kallne1977MeasurementsIons}
Kallne, E., \& Jones, L.~A. 1977, Journal of Physics B Atomic Molecular
  Physics, 10, 3637, \dodoi{10.1088/0022-3700/10/18/020}

\bibitem[{{Kilgus} {et~al.}(1990){Kilgus}, {Berger}, {Blatt}, {Grieser},
  {Habs}, {Hochadel}, {Jaeschke}, {Kr{\"a}mer}, {Neumann}, {Neureither}, {Ott},
  {Schwalm}, {Steck}, {Stokstad}, {Szmola}, {Wolf}, {Schuch}, {M{\"u}ller}, \&
  {Wagner}}]{1990PhRvL..64..737K}
{Kilgus}, G., {Berger}, J., {Blatt}, P., {et~al.} 1990, PRL, 64, 737,
  \dodoi{10.1103/PhysRevLett.64.737}

\bibitem[{Lestinsky {et~al.}(2009)Lestinsky, Badnell, Bernhardt, Grieser,
  Hoffmann, Luki{\'{c}}, M{\"{u}}ller, Orlov, Repnow, Savin, Schmidt, Schnell,
  Schippers, Wolf, \& Yu}]{Lestinsky2009Electron-IonCalculations}
Lestinsky, M., Badnell, N.~R., Bernhardt, D., {et~al.} 2009, ApJ, 698, 648,
  \dodoi{10.1088/0004-637X/698/1/648}

\bibitem[{Linkemann {et~al.}(1995)Linkemann, Kenntner, M{\"{u}}ller, Wolf,
  Habs, Schwalm, Spies, Uwira, Frank, Liedtke, Hofmann, Salzborn, Badnell, \&
  Pindzola}]{Linkemann1995ElectronIons}
Linkemann, J., Kenntner, J., M{\"{u}}ller, A., {et~al.} 1995, Nuclear
  Instruments and Methods in Physics Research B, 98, 154,
  \dodoi{10.1016/0168-583X(95)00095-X}

\bibitem[{Loch {et~al.}(2013)Loch, Pindzola, Ballance, Witthoeft, Foster,
  Smith, \& O'Mullane}]{Loch2013TheModels}
Loch, S., Pindzola, M., Ballance, C., {et~al.} 2013, in American Institute of
  Physics Conference Series, Vol. 1545, American Institute of Physics
  Conference Series, ed. J.~D. Gillaspy, W.~L. Wiese, \& Y.~A. Podpaly,
  242--251, \dodoi{10.1063/1.4815860}

\bibitem[{Loch {et~al.}(2003)Loch, Colgan, Pindzola, Westermann, Scheuermann,
  Aichele, Hathiramani, \& Salzborn}]{Loch2003Electron-impact4}
Loch, S.~D., Colgan, J., Pindzola, M.~S., {et~al.} 2003, Phys. Rev. A, 67,
  42714, \dodoi{10.1103/PhysRevA.67.042714}

\bibitem[{Loewenstein \& Davis(2012)}]{Loewenstein2012An4649}
Loewenstein, M., \& Davis, D.~S. 2012, ApJ, 757, 121,
  \dodoi{10.1088/0004-637X/757/2/121}

\bibitem[{Luki{\'{c}} {et~al.}(2007)Luki{\'{c}}, Schnell, Savin, {Brand}, au,
  Schmidt, B{\"{o}}hm, M{\"{u}}ller, Schippers, Lestinsky, Sprenger, Wolf,
  Altun, \& Badnell}]{Lukic2007DielectronicCalculations}
Luki{\'{c}}, D.~V., Schnell, M., Savin, D.~W., {et~al.} 2007, ApJ, 664, 1244,
  \dodoi{10.1086/519073}

\bibitem[{{Mauche} {et~al.}(2003){Mauche}, {Liedahl}, \&
  {Fournier}}]{2003ApJ...588L.101M}
{Mauche}, C.~W., {Liedahl}, D.~A., \& {Fournier}, K.~B. 2003, \apjl, 588, L101,
  \dodoi{10.1086/375684}

\bibitem[{Mazzotta {et~al.}(1998)Mazzotta, Mazzitelli, Colafrancesco, \&
  Vittorio}]{Mazzotta1998IonizationNI}
Mazzotta, P., Mazzitelli, G., Colafrancesco, S., \& Vittorio, N. 1998, A{\&}AS,
  133, 403, \dodoi{10.1051/aas:1998330}

\bibitem[{{Miller} {et~al.}(2006){Miller}, {Raymond}, {Fabian}, {Steeghs},
  {Homan}, {Reynolds}, {van der Klis}, \& {Wijnands}}]{2006Natur.441..953M}
{Miller}, J.~M., {Raymond}, J., {Fabian}, A., {et~al.} 2006, \nat, 441, 953,
  \dodoi{10.1038/nature04912}

\bibitem[{M{\"{u}}ller(1999)}]{Muller1999PlasmaRings}
M{\"{u}}ller, A. 1999, International Journal of Mass Spectrometry, 192, 9,
  \dodoi{10.1016/S1387-3806(99)00098-6}

\bibitem[{M{\"{u}}ller {et~al.}(2000)M{\"{u}}ller, Teng, Hofmann, Phaneuf, \&
  Salzborn}]{Muller2000AutoionizingIons}
M{\"{u}}ller, A., Teng, H., Hofmann, G., Phaneuf, R.~A., \& Salzborn, E. 2000,
  Phys. Rev. A, 62, 62720, \dodoi{10.1103/PhysRevA.62.062720}

\bibitem[{{Novotn{\'y}} {et~al.}(2012){Novotn{\'y}}, {Badnell}, {Bernhardt},
  {Grieser}, {Hahn}, {Krantz}, {Lestinsky}, {M{\"u}ller}, {Repnow},
  {Schippers}, {Wolf}, \& {Savin}}]{2012ApJ...753...57N}
{Novotn{\'y}}, O., {Badnell}, N.~R., {Bernhardt}, D., {et~al.} 2012, \apj, 753,
  57, \dodoi{10.1088/0004-637X/753/1/57}

\bibitem[{O'Rourke {et~al.}(2001)O'Rourke, Currell, Kuramoto, Li, Ohtani, Tong,
  \& Watanabe}]{ORourke2001Electron-impactIons}
O'Rourke, B., Currell, F.~J., Kuramoto, H., {et~al.} 2001, Journal of Physics B
  Atomic Molecular Physics, 34, 4003, \dodoi{10.1088/0953-4075/34/20/311}

\bibitem[{Pindzola {et~al.}(1986)Pindzola, Griffin, \&
  Bottcher}]{Pindzola1986Electron-impactSequence}
Pindzola, M.~S., Griffin, D.~C., \& Bottcher, C. 1986, Physical Review A, 34,
  3668, \dodoi{10.1103/PhysRevA.34.3668}

\bibitem[{{Pindzola} {et~al.}(1986){Pindzola}, {Griffin}, \&
  {Bottcher}}]{1986PhRvA..34.3668P}
{Pindzola}, M.~S., {Griffin}, D.~C., \& {Bottcher}, C. 1986, PRA, 34, 3668,
  \dodoi{10.1103/PhysRevA.34.3668}

\bibitem[{Raymond \& Smith(1977)}]{Raymond1977SoftPlasma.}
Raymond, J.~C., \& Smith, B.~W. 1977, ApJ, 35, 419, \dodoi{10.1086/190486}

\bibitem[{Rinn {et~al.}(1987)Rinn, Gregory, Wang, Phaneuf, \&
  Mueller}]{Rinn1987Electron-impactMeasurementsb}
Rinn, K., Gregory, D.~C., Wang, L.~J., Phaneuf, R.~A., \& Mueller, A. 1987,
  Physical Review A, 36, 595, \dodoi{10.1103/PhysRevA.36.595}

\bibitem[{Rowan \& Roberts(1979)}]{Rowan1979Electron-impactOxygen}
Rowan, W.~L., \& Roberts, J.~R. 1979, Physical Review A, 19, 90,
  \dodoi{10.1103/PhysRevA.19.90}

\bibitem[{{Savin} {et~al.}(2002){Savin}, {Gwinner}, {Schwalm}, {Wolf},
  {M{\"u}ller}, \& {Schippers}}]{2002nla..work...83S}
{Savin}, D.~W., {Gwinner}, G., {Schwalm}, D., {et~al.} 2002, in NASA Laboratory
  Astrophysics Workshop, ed. F.~{Salama} \& {et al.}, 83

\bibitem[{{Savin} {et~al.}(1997){Savin}, {Kahn}, {Linkemann}, {Saghiri},
  {Schmitt}, {Wolf}, {Bartsch}, {M{\"u}ller}, {Schippers}, \&
  {Chen}}]{1997APS..APR.D1557S}
{Savin}, D.~W., {Kahn}, S.~M., {Linkemann}, J., {et~al.} 1997, in APS April
  Meeting Abstracts, APS Meeting Abstracts, D15.57

\bibitem[{Savin {et~al.}(1999)Savin, Kahn, Linkemann, Saghiri, Schmitt,
  Grieser, Repnow, Schwalm, Wolf, Bartsch, Brandau, Hoffknecht, M{\"{u}}ller,
  Schippers, Chen, \& Badnell}]{Savin1999DielectronicXIX}
Savin, D.~W., Kahn, S.~M., Linkemann, J., {et~al.} 1999, ApJS, 123, 687,
  \dodoi{10.1086/313247}

\bibitem[{{Savin} {et~al.}(1999){Savin}, {Kahn}, {Linkemann}, {Saghiri},
  {Schmitt}, {Grieser}, {Repnow}, {Schwalm}, {Wolf}, {Bartsch}, {Brandau},
  {Hoffknecht}, {M{\"u}ller}, {Schippers}, {Chen}, \&
  {Badnell}}]{1999ApJS..123..687S}
{Savin}, D.~W., {Kahn}, S.~M., {Linkemann}, J., {et~al.} 1999, \apjs, 123, 687,
  \dodoi{10.1086/313247}

\bibitem[{Savin {et~al.}(2002)Savin, Behar, Kahn, Gwinner, Saghiri, Schmitt,
  Grieser, Repnow, Schwalm, Wolf, Bartsch, M{\"{u}}ller, Schippers, Badnell,
  Chen, \& Gorczyca}]{Savin2002DielectronicCalculations}
Savin, D.~W., Behar, E., Kahn, S.~M., {et~al.} 2002, ApJS, 138, 337,
  \dodoi{10.1086/323388}

\bibitem[{Savin {et~al.}(2006)Savin, Gwinner, Grieser, Repnow, Schnell,
  Schwalm, Wolf, Zhou, Kieslich, M{\"{u}}ller, Schippers, Colgan, Loch,
  Badnell, Chen, \& Gu}]{Savin2006DielectronicCalculations}
Savin, D.~W., Gwinner, G., Grieser, M., {et~al.} 2006, ApJ, 642, 1275,
  \dodoi{10.1086/501420}

\bibitem[{{Savin} {et~al.}(2013){Savin}, {Hahn}, {Becker}, {Bernhardt},
  {Grieser}, {Krantz}, {Lestinksy}, {Mueller}, {Novotny}, {Repnow},
  {Schippers}, {Spruck}, \& {Wolf}}]{2013HEAD...1312201S}
{Savin}, D.~W., {Hahn}, M., {Becker}, A., {et~al.} 2013, in AAS/High Energy
  Astrophysics Division \#13, AAS/High Energy Astrophysics Division, 122.01

\bibitem[{Schippers {et~al.}(2010)Schippers, Lestinsky, M{\"{u}}ller, Savin,
  Schmidt, \& Wolf}]{Schippers2010DielectronicExperiments}
Schippers, S., Lestinsky, M., M{\"{u}}ller, A., {et~al.} 2010, arXiv e-prints,
  arXiv:1002.3678

\bibitem[{Schmidt {et~al.}(2006)Schmidt, Schippers, M{\"{u}}ller, Lestinsky,
  Sprenger, Grieser, Repnow, Wolf, Brandau, Luki{\'{c}}, Schnell, \&
  Savin}]{Schmidt2006Electron-IonXIII}
Schmidt, E.~W., Schippers, S., M{\"{u}}ller, A., {et~al.} 2006, A{\&}A, 641,
  L157, \dodoi{10.1086/504038}

\bibitem[{{Schmidt} {et~al.}(2008){Schmidt}, {Schippers}, {Bernhardt},
  {M{\"u}ller}, {Hoffmann}, {Lestinsky}, {Orlov}, {Wolf}, {Luki{\'c}}, {Savin},
  \& {Badnell}}]{2008AA...492..265S}
{Schmidt}, E.~W., {Schippers}, S., {Bernhardt}, D., {et~al.} 2008, \aap, 492,
  265, \dodoi{10.1051/0004-6361:200810834}

\bibitem[{Schmidt {et~al.}(2009)Schmidt, Bernhardt, Hoffmann, Lestinsky,
  Luki{\'{c}}, M{\"{u}}ller, Orlov, Savin, Schippers, \&
  Wolf}]{Schmidt2009ExperimentalIron}
Schmidt, E.~W., Bernhardt, D., Hoffmann, J., {et~al.} 2009, in Journal of
  Physics Conference Series, Vol. 163, Journal of Physics Conference Series,
  12028, \dodoi{10.1088/1742-6596/163/1/012028}

\bibitem[{Smith \& Brickhouse(2014)}]{Smith2014ChapterPlasmas}
Smith, R.~K., \& Brickhouse, N.~S. 2014, Advances in Atomic Molecular and
  Optical Physics, 63, 271, \dodoi{10.1016/B978-0-12-800129-5.00004-3}

\bibitem[{Smith {et~al.}(2001)Smith, Brickhouse, Liedahl, \&
  Raymond}]{Smith2001CollisionalIonsb}
Smith, R.~K., Brickhouse, N.~S., Liedahl, D.~A., \& Raymond, J.~C. 2001, ApJL,
  556, L91, \dodoi{10.1086/322992}

\bibitem[{{Summers}(1974)}]{1974MNRAS.169..663S}
{Summers}, H.~P. 1974, \mnras, 169, 663, \dodoi{10.1093/mnras/169.3.663}

\bibitem[{Sutherland \& Dopita(1993)}]{Sutherland1993CoolingPlasmas}
Sutherland, R.~S., \& Dopita, M.~A. 1993, ApJ, 88, 253, \dodoi{10.1086/191823}

\bibitem[{Urdampilleta {et~al.}(2017)Urdampilleta, Kaastra, \&
  Mehdipour}]{Urdampilleta2017X-rayZn}
Urdampilleta, I., Kaastra, J.~S., \& Mehdipour, M. 2017, A{\&}A, 601, A85,
  \dodoi{10.1051/0004-6361/201630170}

\bibitem[{Wang {et~al.}(1988)Wang, Griem, Hess, \&
  Rowan}]{Wang1988MeasurementPlasmas}
Wang, J.-S., Griem, H.~R., Hess, R., \& Rowan, W.~L. 1988, Physical Review A,
  38, 4761, \dodoi{10.1103/PhysRevA.38.4761}

\bibitem[{Wong {et~al.}(1993)Wong, Beiersdorfer, Chen, Marrs, Reed, Scofield,
  Vogel, \& Zasadzinski}]{Wong1993Electron-impactFe23+}
Wong, K.~L., Beiersdorfer, P., Chen, M.~H., {et~al.} 1993, Phys. Rev. A, 48,
  2850, \dodoi{10.1103/PhysRevA.48.2850}

\bibitem[{{Younger}(1981)}]{1981PhRvA..24.1278Y}
{Younger}, S.~M. 1981, \pra, 24, 1278, \dodoi{10.1103/PhysRevA.24.1278}

\bibitem[{{Zatsarinny} {et~al.}(2004{\natexlab{a}}){Zatsarinny}, {Gorczyca},
  {Korista}, {Badnell}, \& {Savin}}]{Zatsarinny2004DRNeon}
{Zatsarinny}, O., {Gorczyca}, T.~W., {Korista}, K., {Badnell}, N.~R., \&
  {Savin}, D.~W. 2004{\natexlab{a}}, A\&A, 426, 699,
  \dodoi{10.1051/0004-6361:20040463}

\bibitem[{{Zatsarinny} {et~al.}(2004{\natexlab{b}}){Zatsarinny}, {Gorczyca},
  {Korista}, {Badnell}, \& {Savin}}]{Zatsarinny2004DRCarbon}
{Zatsarinny}, O., {Gorczyca}, T.~W., {Korista}, K.~T., {Badnell}, N.~R., \&
  {Savin}, D.~W. 2004{\natexlab{b}}, \aap, 417, 1173,
  \dodoi{10.1051/0004-6361:20034174}

\bibitem[{Zhang \& Sampson(1990)}]{Zhang1990RapidIons}
Zhang, H.~L., \& Sampson, D.~H. 1990, Physical Review A, 42, 5378,
  \dodoi{10.1103/PhysRevA.42.5378}

\end{thebibliography}

\rr{\appendix}

\rr{
The tools and CSD error estimates used for this paper are available on the GitHub distribution. \texttt{variableapec} requires the \texttt{pyatomdb} package and Python 3. Below are examples of how to access the CSD error estimates we have completed and reproduce some of the figures in this paper using routines that can be used for other elements and lines.} 

\vspace{5mm}
The systematic uncertainties we have gathered are listed in a dictionary with the element number Z and a dictionary of ionization and recombination errors for each ion as key-value pairs. Currently we have only completed CSD error estimates for O and Fe. We welcome any contributions to add errors for other ions and elements. 

\vspace{1mm}
\texttt{>>> oxygen = variableapec.systematics[8]}

\texttt{>>> iron = variableapec.systematics[26]}

\vspace{5mm}
To perturb all rate coefficients via Monte Carlo calculations like Figure \ref{fig:new_csd}, supply the element number, ionization and recombination errors,  and the number of runs (default is 100): 

\vspace{1mm}
\texttt{>>> Z, max\_ionize, max\_recomb = 26, iron[`ionize'], iron[`recomb']}

\texttt{>>> variableapec.monte\_carlo\_csd(Z, max\_ionize, max\_recomb, runs=500)}

\vspace{5mm}
To compare the rate of change in ionic concentration from varying a single rate coefficient with multiple perturbation magnitudes like Figure \ref{fig:slope}, supply the element, ion, type of rate coefficient to vary, and a list of positive rate errors. 

\vspace{1mm}
\texttt{>>> Z, z1, rate\_type, errors = 8, 7, `i', [0.05, 0.1, 0.15, 0.2]} 

\texttt{>>> variableapec.csd\_slope(Z, z1, rate\_type, errors)}

\vspace{5mm}
To perturb all rate coefficients via Monte Carlo calculations and calculate the change in peak ionic concentration as a function of error on rate coefficients like Figure \ref{fig:O_abund} [left], supply the element and the fractional errors to perturb all rate coefficients by. If a z1\_interesting is given, the routine will create a second plot from the same Monte Carlo calculations of the change in ionic concentration of that ion over temperatures where the unperturbed concentration is greater than $10^{-3}$\ (see Figure \ref{fig:O_abund} [right]). If no z1\_interesting given, only the change in peak ionic concentration will be plotted for z1\_plot, which is a list of all ions by default:

\vspace{1mm}
\texttt{>>> Z, errors, z1\_interesting = 8, [0.05, 0.1, 0.15, 0.2, 0.25, 0.3], 6} 

\texttt{>>> variableapec.ion\_sensitivity(Z, errors, z1\_interesting=z1\_interesting)}

\vspace{5mm}
To check the sensitivity of a line to a perturbation in the direct excitation rate or $A$ value of another transition, supply the element, ion, sensitive transition, rate type and rate errors, the transition to be varied, and the temperatures or densities to look at. The default density is 1 cm$^{-3}$ and the default temperatures are the five temperatures of interest [see Figure \ref{fig:5_temps}]. 

\vspace{5mm}
To check the sensitivity due to only changes in direct excitation rate like Figure \ref{fig:Fe_line_sens} [left]:

\vspace{1mm}
\texttt{>>> Z, z1, up, lo, rate\_type, rate\_errors, transition\_list = 26, 17, 17, 6, `exc', [0.05, 0.1, 0.15, 0.2, 0.25, 0.3], [(17,1)]}

\texttt{>>> variableapec.line\_sensitivity(Z, z1, up, lo, rate\_type, rate\_error, transition\_list)}

\vspace{5mm}
To check the sensitivity to changes in direct excitation rate in addition to CSD errors like Figure \ref{fig:Fe_line_sens} [right]:

\vspace{1mm}
\texttt{>>> Z, z1, up, lo, rate\_type, rate\_errors, transition\_list, max\_ionize, max\_recomb = 26, 17, 17, 6, `exc', [0.05, 0.1, 0.15, 0.2, 0.25, 0.3], [(17,1)], iron[`max\_ionize'], iron[`max\_recomb']}

\texttt{>>> variableapec.line\_sensitivity\_csd(Z, z1, up, lo, rate\_type, errors, transition\_list, max\_ionize, max\_recomb)}

\end{document}